\documentclass[10pt,conference]{IEEEtran}
\AtBeginDocument{%
  \providecommand\BibTeX{{%
    \normalfont B\kern-0.5em{\scshape i\kern-0.25em b}\kern-0.8em\TeX}}}

\usepackage[framemethod=tikz]{mdframed} % 
\usepackage{enumitem}
\usepackage{soul}
\usepackage{stfloats}
\usepackage[numbers]{natbib}
\usepackage[hidelinks]{hyperref}

\usepackage{soul}
\usepackage{colortbl}
\usepackage{adjustbox}
\usepackage{multirow}
\usepackage{tikz}
\usepackage{xcolor}
\usepackage{refcount}
\usepackage{hyperref}
\usepackage{tabularx}
\usepackage{threeparttable}
\usepackage{multicol}
\usepackage{float}
\usepackage{siunitx}
\usepackage{booktabs}

\usepackage[normalem]{ulem}
\newcommand{\boldification}[1]{\ifdraft\indent **\textbf{#1}**\\\indent\else\relax\fi}

\definecolor{Cream}{RGB}{254, 238, 214}

\newif\ifdraft
\draftfalse %hides boldifications and comments
\newif\ifrevising
   \revisingfalse %makes revisions normal appearance
\newcommand{\revised}[1]{{\ifrevising{\color{blue}#1}\else{\color{black}#1}\fi}}
 \newcommand{\deleted}[1]{{\ifrevising{\relax}\else\relax\fi}}

\begin{document}

% \title{Tales from Trenches: Student Experiences on Using GenAI for Learning Software Engineering}

% \title{Where, when, and How to incorporate GenAI in Education: reflections from the trenches}
% \title{Where and When GenAI helps SE students: Reflections from the trenches}

%\title{Where to (not) use it? Unveiling the Perils of Generative AI for Software Engineering Students}

% \title{On Becoming the Future Generation of AI-Savvy Software Engineers: A View From the Trenches }
 \title{Insights from the Frontline: GenAI Utilization Among Software Engineering Students}
%\title{From Novices to AI-Savvy Software Engineers: Reflections of Students' GenAI Usage Journey}

% \author{\IEEEauthorblockN{Anonymous Authors}}

\vspace{-4mm}
\author{
\IEEEauthorblockN{
Rudrajit Choudhuri\IEEEauthorrefmark{1}, Ambareesh Ramakrishnan\IEEEauthorrefmark{1}, Amreeta Chatterjee\IEEEauthorrefmark{1}, Bianca Trinkenreich\IEEEauthorrefmark{1},\\ Igor Steinmacher\IEEEauthorrefmark{2}, Marco Gerosa\IEEEauthorrefmark{2}, Anita Sarma\IEEEauthorrefmark{1}
	\\
	}
\\
\vspace{-4mm}
\IEEEauthorblockA{\IEEEauthorrefmark{1}Oregon State University, United States, \{choudhru, ramakria, chattera, bianca.trinkenreich, anita.sarma\}@oregonstate.edu}
\IEEEauthorblockA{\IEEEauthorrefmark{2}Northern Arizona University, United States, \{igor.steinmacher, marco.gerosa\}@nau.edu}
\vspace{-9mm}
}

\newcommand{\explaintwo}[1]{%
\par%
\noindent\fbox{%
    \parbox{\dimexpr\linewidth-2\fboxsep-2\fboxrule}{#1}%
}%
}

% \newcounter{guidelineno}
%\newcommand{\guideline}[1]{\noindent\textcolor{blue}{\textbf{DG \refstepcounter{guidelineno}\theguidelineno:} #1}}

%\newcounter{fullguidelineno}
% \newcommand{\fullguideline}[2]{\noindent\textcolor{blue}{\textbf{DG \refstepcounter{guidelineno}\theguidelineno: #1} #2}}
\newcommand{\blue}[1]{\textcolor{blue}{#1}}

%\newcommand{\guideline}[1]{}

% Whats the prob: Educators speculating and want proper landscape of where genAI helps/hinders. 

% Why is a prob: Students are using it adhoc, leading to problemssss

% What others did: Pick out benefits and challenges

% What we do: (1) Landscape, (2) associations

% Why you care? Clear picture of where it helps/hinders and why/what causes and what consequences.
\maketitle
\IEEEpeerreviewmaketitle

\begin{abstract}
Generative AI (genAI) tools (e.g., ChatGPT, Copilot) have become ubiquitous in software engineering (SE). As SE educators, it behooves us to understand the consequences of genAI usage among SE students and to create a holistic view of where these tools can be successfully used. 
Through 16 reflective interviews with SE students, we explored their academic experiences of using genAI tools to complement SE learning and implementations. We uncover the contexts where these tools are helpful and where they pose challenges, along with examining why these challenges arise and how they impact students. We validated our findings through member checking and triangulation with instructors. Our findings provide practical considerations of where and why genAI should (not) be used in the context of supporting SE students.
\end{abstract}

%%
%% The code below is generated by the tool at http://dl.acm.org/ccs.cfm.
%% Please copy and paste the code instead of the example below.
%%
% \begin{CCSXML}
% <ccs2012>
% <concept>
% <concept_id>10003120.10003121.10011748</concept_id>
% <concept_desc>Human-centered computing~Empirical studies in HCI</concept_desc>
% <concept_significance>500</concept_significance>
% </concept>
% </ccs2012>
% \end{CCSXML}

% \ccsdesc[500]{Human-centered computing~Empirical studies in HCI}
% %%
%% Keywords. The author(s) should pick words that accurately describe
%% the work being presented. Separate the keywords with commas.
% \keywords{Empirical Study, Generative AI, Challenges, Software Engineering}

\begin{IEEEkeywords}
Empirical Study, Generative AI, Challenges and Benefits, Software Engineering Education, Human-AI Interaction
\end{IEEEkeywords}

%%
%% This command processes the author and affiliation and title
%% information and builds the first part of the formatted document.

\vspace{-3mm}
% Blog link: https://medium.com/bits-and-behavior/more-than-calculators-why-large-language-models-threaten-public-education-480dd5300939
% \vspace{-2mm}
\section{Introduction}
\label{sec:intro}

Generative AI (genAI) tools (e.g., ChatGPT \cite{GPT4}, Gemini \cite{Gemini}, Copilot \cite{Copilot}) have asserted their prevalence over a couple of years. These tools are widely used in software development as well as education \cite{fan2023large, ebert2023generative, russo2023navigating}. Thus, it behooves us as educators to understand how and where to incorporate genAI to train the next generation of AI-savvy SE workforce \cite{bull2023generative, nguyen2023generative}.

% \boldification{Given the prevalence of this technology, there is a hot debate as to if students should use genAI when they learn. Some say this is the end of education and the workforce we create will not know how to code, others say its a useful tool like the calculator**}
However, how to seamlessly incorporate genAI use by SE students without long-lasting adverse effects is a non-trivial question.
There are intense debates about the appropriateness of students using genAI in their learning processes \cite{denny2024computing}. Some argue that this marks the end of traditional education \cite{welsh2022end, yellin2023premature} and fear that the future workforce will lack essential skills, such as problem-solving, due to over-reliance on AI tools \cite{chen2021evaluating, bommasani2021opportunities}. The counterargument is that it is not all negative; there are opportunities despite the risks \cite{amoozadeh2023towards, becker2023programming, denny2024computing,penney2023conversations} and ``it depends'' on how these tools are being used.

% \boldification{the concerns of the former group are valid because....(1) Providing convincing evidence often leads to (over)reliance. These tools are trained off the internet. What happens when LLMs are wrong? Students using it for learning are not capable of judging the extent of accuracy.}
 
The concerns raised by the former group are serious. First, genAI systems are not infallible. They are trained on available (sometimes biased or incomplete) data, and function by predicting the most likely sequence of symbols, without guaranteeing the veracity of information \cite{vaswani2017attention, achiam2023gpt}. The convincing nature of genAI responses, even when it is hallucinating \cite{choudhuri2024far}, can foster over-reliance among novices \cite{chen2021evaluating, bommasani2021opportunities}, who might not have the relevant knowledge to assess the accuracy of its responses. 
% \boldification{** (2) It can take away from learning for students who want to bypass learning and find the shortest path and for students who actually want to learn. Why? Comprehending a solution fetched from genAI is not the same as coming up with one. Problem solving boiled down to a reading comprehension.}
Second, a deeper concern is that genAI will stymie learning: not only for the ones who want to bypass learning, seeking ready-made answers instead of engaging with the material; but also for the ones who actually want to learn. This is because of the fundamental difference between comprehending a solution provided by genAI and creating a solution through personal, mental effort. Using genAI reduces students' problem-solving and critical thinking into a mere reading comprehension exercise \cite{ko_2024}, undermining metacognition, the very essence of the learning process.

% \boldification{**but, the genie is already out of the lamp...and its not feasible to forbid students to use genAI. Yes, some will cheat, but how can we help students use LLM as a tool to be more efficient.}
% \boldification{Additionally, Industry is all hyped up about LLMs, so, having experience with LLMs will help students get better jobs and more importantly know how to utilize it well}
% \boldification{We are not alone in this thinking. Educators, policy makers, are currently speculating of genAI incorporation into curricula/pedagogy and want a clear landscape of where it can help/hinder.}
% 
Yet, considering a complete ``ban" or forbidding students from using genAI tools is impractical \cite{lau2023ban}, as these tools are already widely adopted \cite{russo2023navigating, liang2023understanding}. 
% Additionally, banning these tools can potentially create of creating an AI digital divide \cite{carter2020exploring}, marginalizing those without access or skills to use these technologies. 
Further, with the software industry's exuberant adoption of genAI, proficiency with genAI has become a necessary job skill for the future workforce \cite{whiting_2023, woodruff2023knowledge}. Acknowledging the inevitability of some instances of academic misconduct \cite{prather2023robots}, the emphasis should thus shift towards guiding students on the effective use of genAI. Educators and policymakers aligned with this thought are speculating on the incorporation of genAI into curricula and pedagogy in the long term \cite{lau2023ban, malik2023so, bull2023generative}, seeking a clear landscape of where these tools can help or hinder.
%
% \boldification{Current work has looked into benefits and challenges from students’ and educators’ perspectives highlighting the “what's” in the picture...but there is a gap in understanding the landscape of use along with the “where” and “why’s” behind students’ struggles. **this gap makes the world end }
% %
Prior research has explored the potential benefits and challenges of genAI usage from both student \cite{rajabi2023exploring, joshi2023let, raman2023university} and educator \cite{lau2023ban, rajabi2023exploring, wang2023towards, zastudil2023generative} perspectives. %, identifying the ``what's" in the picture. 

However, there is a dearth in understanding the current state of genAI usage among SE students: \textit{when} and \textit{how} they use genAI, \textit{where} they are finding benefits or facing challenges, and \textit{why} these challenges exist. Without answers to these questions, we will not know how to effectively integrate genAI into (SE) education without risking adverse impacts.

%Closing this gap is essential for offering insights into effective genAI integration for enhancing students' learning experiences, complimenting rather than complicating existing approaches.

%—identifying ``where" they succeed or struggle, along with explorations into the ``why" aspects of these challenges. 

\boldification{** Our work, closes this gap, looks into ….. RQ 1 and RQ 2.}
In this paper, we investigate: \textbf{\textit{(RQ1): What is the current state of genAI usage among SE students?}} Through reflective interviews with 16 SE students, validated through member checking and interviews with two instructors, we elicit students' experiences of genAI usage to complement learning of SE concepts and their implementations in SE courses. We explored the specific contexts when students turn to genAI tools, examining how they incorporate these tools in their work, and where they face benefits and challenges. 

Furthermore, to gain a deeper understanding of the appropriateness of genAI tools for supporting SE students, it is essential to identify why students face challenges when using these tools. This led to our second research question: \textbf{\textit{(RQ2): What are the causes and consequences of the challenges faced by SE students?}} 
Qualitative analysis of student experiences revealed six intrinsic categories of issues (faults and gaps) within genAI itself, which contributed to five categories of student challenges. These challenges ultimately impacted students' learning and task outcomes, self-perception, and willingness to adopt genAI technology. 
% Further, we discuss the nuances in these associations, highlighting their differences across learning and implementation phases.

% \boldification{** providing practical considerations of where genAI can help/hinder by providing the landscape of genAI usage in the SE context, investigating the (when, why, and how students use genAI, where they face benefits and challenges) along with explaining the causes and consequences of the struggle (arising from AI deficiencies and leading to human impacts).} 

% \boldification{** Educators should caution students of clear cases where genAI fails, and show how to best use genAI. where and to what extent it can support students in SE learning, as the concepts for SE are beyond just text book learning, it extends into the workplace, where software developers need to learn job-relevant topics, processes, and tools.} 

The primary contributions of this paper are twofold: (1) It assesses the use of genAI among SE students, detailing the circumstances (``when") and methods (``how") of their genAI usage. This evaluation further explores the contexts (``where") students perceive benefits and challenges, offering practical insights for educators interested in incorporating genAI to enhance student learning experiences. (2) It identifies the causes and consequences of the challenges (``why's") faced by students in using genAI. These findings offer practical insights for educators to guide students on where and why genAI can (or cannot) be effectively used in the context of SE education.

% These findings will aid educators in guiding students toward effective genAI use in SE education.
%and the adversities of using the technology in learning contexts.

% =====END of Intro======
% ** A recent blog drew a parallel between the era of calculators and the age of LLMs. - Amy Ko.

% ** The posed threat that calculators would take away from math education was mostly a false alarm, because even if calculators could present numerical answers, it couldn’t explain how it got to it.

% ** Calculators are deterministic systems, LLMs are probabilistic. LLMs too can’t provide correct explanations, but they provide more than numerical answers: convincing and often misleading “likely sequence of symbols” – raising bigger concerns and leaving educators in tech mess that they didn't ask for.

% \vspace{-3mm}
\vspace{-2mm}
\section{Related Work}
\label{sec:background}

\textbf{User studies with genAI}: Research on genAI-based programming tools \cite{barke2023grounded, kazemitabaar2023studying, mozannar2022reading, ross2023programmer, vaithilingam2022expectation, xu2022ide, ziegler2022productivity, prather2023s} has shown their usefulness and limitations. Vaithilingam et al. \cite{vaithilingam2022expectation} found programmers faced more difficulties completing tasks with GitHub Copilot compared to traditional autocomplete, although with no significant effects on task completion time. 
Barke et al. \cite{barke2023grounded} investigated programmers' interaction with Copilot identifying exploration (using genAI for planning) and acceleration (using genAI to speed up code authoring) interaction modes. They observed that over-reliance on Copilot can lead to increased cognitive load. Bird et al. \cite{bird2023taking} studied how developers who were first-time Copilot users, engaged with it. They found that developers accepted generated suggestions for efficiency, trading autonomy and control over code. Other studies \cite{liang2023understanding, ziegler2022productivity} focused on programmers' perceptions, with Liang et al. \cite{liang2023understanding} finding appreciation for autocomplete features but concerns over code quality, control, and IP risks.
While these studies offer insights into user perception of genAI tools, their findings primarily apply to professional programmers. Closest to our work is the study by Choudhuri et al. \cite{choudhuri2024far}. They investigated the effectiveness and pitfalls of ChatGPT in supporting students in SE tasks. Their findings highlight that using ChatGPT had no significant improvements in productivity, although it significantly increased participants’ frustration. Further, they investigated the causes and consequences of AI faults arising from human-AI interaction guideline \cite{amershi2019guidelines} violations, leading to negative consequences for students using ChatGPT for SE implementations. 

\textsc{\textbf{(+)}} In contrast to these studies, our study did not control or focus on specific tasks or genAI tools. We complement their findings by concentrating on student challenges and presenting associations between their causes and implications across the SE learning and implementation phases.

\textbf{GenAI in education}: Previous research has examined the use of genAI tools in education, with research investigating both the opportunities and challenges these AI tools present \cite{
amoozadeh2023towards, becker2023programming, denny2024computing, lau2023ban, wang2023towards, prather2024widening, skripchuk2024investigation}. This included advantages, impact on teaching methods due to readily available solutions, and concerns over plagiarism, biases, and fostering detrimental habits among students. In work exploring the opportunities and risks presented by these tools, Bommasani et al.\cite{bommasani2021opportunities} explicitly list Copilot as a challenge for educators, stating students' over-reliance may negatively impact their learning. How students should adopt genAI tools remains unclear \cite{ernst2022ai}, but their increasing role inside and outside the classroom seems certain. Beyond theoretical discussion, practical research efforts have aimed at enhancing student assistance in code generation on introductory programming problems \cite{becker2019compiler}, providing learning resources \cite{finnie2022robots, sarsa2022automatic, savelka2023can}, explanations \cite{leinonen2023comparing, macneil2023experiences, sarsa2022automatic, wermelinger2023using}, and finding issues in problematic code \cite{leinonen2023using}. While suggesting significant aid, these studies highlight varying extents of help based on task complexity and prompt quality, emphasizing the need for clarification on where genAI can be effectively used.

\textsc{\textbf{(+)}} Our study adds to this body of work by presenting the current state of genAI usage among SE students (when and how), alongside where students perceive benefits and challenges, giving practical considerations for educators to integrate genAI in SE education. 

% References in student and instructor perspectives from this paper: https://dl.acm.org/doi/pdf/10.1145/3623762.3633499 

\textbf{Student and instructor perspectives}: Recent studies have examined student and instructor perceptions of genAI tools in broad educational settings, revealing a generation gap in perceptions \cite{chan2023ai, padiyath2024insights}. Students are generally optimistic, but are self-aware of the negative impacts of AI on learning \cite{padiyath2024insights}, while instructors express concerns about over-reliance, ethics, and skepticism regarding genAI's abilities \cite{lau2023ban, rajabi2023exploring, raman2023university, wang2023towards, zastudil2023generative}. Instructors emphasize the urgent need for clear policies and guidelines on where and to what extent genAI should be integrated to maintain academic integrity and equitable learning. Lau and Guo \cite{lau2023ban} interviewed instructors about adapting genAI tools like ChatGPT and Copilot in courseware, framed around a hypothetical futuristic scenario. Instructors' short-term concerns centered on cheating, plagiarism, and reliance on invigilated exams, while long-term perspectives varied between resisting AI tools and integrating them into the curriculum. 
% The interviews were framed around a hypothetical scenario where students had access to an AI tool capable of writing perfect code for any programming problem. 
Zastudil et al.’s \cite{zastudil2023generative} comparative analysis showed aligned concerns between students and instructors regarding over-reliance, trustworthiness, and plagiarism, but divergent preferences for addressing these issues. Students focused on the quality of genAI responses, whereas instructors worried about students' ability to identify incorrect or misleading responses. 
% Notably, Joshi et al. [27] conducted a student-centric investigation, focusing on usage patterns, benefits, and challenges of ChatGPT among undergraduates, highlighting usability, reliability, learning, and ethical challenges
Other student-centric investigations \cite{joshi2023let, kim2024chatgpt} have explored students' querying behavior, types of assistance sought, as well as benefits and challenges, emphasizing concerns related to usability, reliability, engagement, learning, and ethics.

\textbf{(+)} Prior work has identified the ``what's" --- benefits, challenges, and perspectives --- from both student and instructor viewpoints. Our work analyzes the “where” and “why”, contributing an understanding of the current genAI usage in SE education, alongside explaining the causes and consequences of challenges faced by SE students in using these tools.

% \vspace{-3mm}
\vspace{-1mm}
\section{Method}
 
We conducted semi-structured interviews with software engineering (SE) students to elicit their academic experiences with genAI. See supplemental \cite{supplemental2023} for the interview script.

\vspace{-1mm}
\subsection{Interview Planning}
\textbf{Script design}: 
Learning is a continuous process that begins with acquiring and understanding knowledge and progresses to applying and synthesizing that knowledge in new contexts \cite{krathwohl2002revision, forehand2005bloom}. In the context of SE education, this spans from learning SE concepts to their implementations \cite{cabezas2020using}. To understand how SE students use genAI to complement their SE learning, we structured our interviews into three parts. The first two parts focused on students' experiences in using genAI to complement their (1) learning of SE concepts and (2) implementation of SE tasks. The final part focused on their (3) overall experiences of using genAI tools.

In the first two parts, we asked questions about (1) their reasons for using genAI tools, (2) how and in what contexts they leveraged these tools, and (3) the benefits and challenges they perceived, along with the specific contexts and reasons for these perceptions. We anchored these discussions in tangible artifacts, as suggested by \citet{lau2023ban}, by requesting participants to examine their conversation history with genAI tools (e.g., ChatGPT) and reflect on their experiences through these concrete examples. Participants' previous conversation threads with genAI acted as a foundation to support deeper recollections of their experiences. 
A potential risk of focusing on specific artifacts is that participants can fixate on specific conversations or experiences. Therefore, to balance detail-oriented and holistic perspectives, we structured the final part of our script to gather participants' overall experiences with genAI. In the end, a wrap-up question invited any additional insights and feedback not covered by the structured questions.
% , offering participants an open-ended opportunity to share insights or experiences 

To reduce cognitive biases such as priming \cite{bargh2000mind} and anchoring \cite{furnham2011literature}, we intentionally avoided mentioning specific tools, such as ChatGPT, in our interview protocol. Everything was worded as ``genAI tools''. However, if participants initiated discussions about particular tools, we allowed the conversation to naturally shift toward those tools. Before the main interviews, we conducted seven sandbox sessions and a pilot to test and refine our interview scripts. This led to simplifying our script for clearer, more straightforward questions. The results reported in this paper are based on the actual 16 interviews, excluding these preliminary sessions. We did not segregate the interviews and data analysis into distinct phases. Instead, we continuously refined our understanding, adjusting our interview script and code set as necessary, in line with recommended practices in qualitative research \cite{creswell2016qualitative}.
% The core interview questions were stable across all interviews. 

% {CS492: Mobile Software Development, CS494: Advanced Web Development)}
% {CS462: Software Engineering Project, CS467: Senior Software Engineering Capstone}
\textbf{Recruitment}: Our goal was to elicit students' experiences using genAI to complement their SE learning and implementations. So, we conferred with SE instructors at our university, who recommended senior SE courses and SE capstone courses,
%\footnote{CS492, CS494. Course names anonymized for review} and SE capstone courses\footnote{CS462, CS467}, 
as students from these classes would have had the opportunity to use genAI to learn concepts and help implement solutions. Our selection criteria were students who had: (a) familiarity with genAI and experience with it in SE, (b) completed both introductory and intermediate SE courses in the previous year (2023), 
% CS361, CS362, CS492, CS494
and (c) enrolled in senior SE or capstone courses. Instructors facilitated recruitment by announcing the study in both in-person and e-campus classes. Interested students completed a questionnaire about their demographics (age, gender, and academic level), familiarity with genAI, and experience with genAI in SE. We received responses from 33 students, of which 28 satisfied the selection criteria.

\vspace{-2mm}
\subsection{Data Collection and Analysis}
% \vspace{-1mm}

Interviews were conducted remotely through Zoom, adhering to the university's IRB protocol. Before their participation, participants agreed to an IRB-approved informed consent. The sessions were audio-recorded and lasted around an hour each. The first author transcribed the interviews.

\textbf{Participants:} We initially invited 28 students who met our selection criteria, inquiring about their availability for participation. Out of these, 19 confirmed their availability. However, three students withdrew, leading to a final count of 16 participants: 10 self-identified as men, five as women, and one as non-binary or gender diverse. In subsequent sections, participants are denoted as P1-P16. Their demographics are summarized in Table \ref{table:demographics}. As a token of appreciation, students received a \$20 Amazon gift card. 

% https://docs.google.com/spreadsheets/d/14nAXsufAZltljnFZ5XbIUG_Y7_VSltF6FP7GOE4PKVs/edit#gid=0

\begin{table}[!th]
\centering
\vspace{-4mm}
\caption{Participant Demographics (n=16)}
\label{tab:demographics}
\scriptsize
\vspace{-3mm}
\begin{tabular}{p{5cm}
                S[table-format=4.0]
                S[table-format=2.1]}
\toprule
\textbf{\textit{Attribute}} & {\textbf{\textit{N}}} & {\textbf{\textit{Percentage}}}\\
\midrule
\midrule

\multicolumn{3}{c}{Gender}\\
\midrule

Men & 10 & 62.5\%\\
Women & 5 & 31.3\%\\
Non-Binary or Gender Diverse & 1 & 6.3\%\\
\midrule

\multicolumn{3}{c}{Academic Level}\\
\midrule
Junior & 4 & 25.0\%\\
Senior & 12 & 75.0\%\\

\midrule

\multicolumn{3}{c}{GenAI Used}\\
\midrule
ChatGPT & 16 & 100.0\%\\
Gemini/Bard & 5 & 31.2\%\\
Copilot & 4 & 25.0\%\\
% ChatGPT, Gemini/Bard & 4 & 25.0\%\\
% ChatGPT, Copilot, \& Gemini/Bard & 1 & 6.3\%\\

% \midrule

% \rowcolor{gray!30}\multicolumn{3}{c}{Usage Frequency}\\
% Medium & 3 & 18.7\%\\
% High & 13 & 81.3\%\\

\bottomrule

\end{tabular}
\label{table:demographics}
\vspace{-1mm}
\end{table}

\textbf{Qualitative analysis}: We analyzed the data using \textit{reflexive thematic analysis} \cite{braun2006using, braun2022conceptual} to discern patterns and meanings within it. We used Atlas.ti \cite{atlas_website} to facilitate this process, which involved an iterative method of adjusting codes, e.g., merging and splitting them, as our understanding of the data deepened and themes began to emerge. To ensure reliability in our analysis, we held multiple team meetings over seven weeks. During these sessions, the authors critically compared and contrasted the codes and discussed the differences, as advocated for in thematic analysis \cite{creswell2016qualitative, mcdonald2019reliability}. Our themes are an \textit{output} of the analysis process rather than an input, as occurs in other forms of qualitative analysis. In more detail, we proceeded as follows:

Two authors inductively open-coded the data, identifying preliminary codes. Subsequent team meetings were dedicated to refining these codes as a shared understanding of the data developed. As the analysis progressed, the authors built post-formed codes and associated them with respective parts of the interview transcripts, following a negotiated agreement. Iterative adjustments to the code set and the interview script were made as necessary.
% 
% We then performed a second pass of axial coding, comparing and contrasting codes and merging (or splitting) them as required. 
% 
Next, we compared and contrasted the codes, merging (or splitting) them as required. A key aspect of this step was determining the granularity of the codes, which was achieved by analyzing co-occurrences: codes that frequently appeared together were merged; otherwise, they remained separate. During this process, codes with logical connections were grouped into higher-level categories.

%Themes emerged and were continuously refined until team consensus was reached. We conducted a comprehensive review of the themes against the entire dataset, allowing for the refinement and reorganization of codes to better reflect the data. We analyzed in tandem with data collection, reaching a point of convergence during the writing process with links formed between the research questions, the existing literature, and the empirical data.

We reached saturation after conducting 10 interviews, beyond which subsequent data did not yield new insights \cite{francis2010adequate}. Nonetheless, we proceeded with six additional interviews to ensure that no other insights emerged, as well as to focus on achieving a balanced representation of participant demographics and incorporating diverse perspectives.

We validated our findings through member checking with interviewees and then triangulated the findings with (1) social science theories and (2) interviews with instructors from the courses where we recruited participants.

1) \textbf{Member checking with students}: 
Following data analysis, we % performed member checking to validate our findings. We 
emailed a questionnaire to our participants, presenting each of our findings with a Likert scale for agreement and an open-ended text option for feedback (or corrections). All interviewees except P10 and P14 responded. Participant feedback confirmed our findings, with some providing additional clarifications, but no new insights or disagreements emerged.

2) \revised{\textbf{Instructor interviews}: We interviewed SE instructors, presenting to them where students reported the benefits and challenges of using genAI along with the consequences of these challenges. The instructors confirmed all findings, with one exception, as noted in the results (see supplemental for detailed examples). The interview script is available in the supplemental material \cite{supplemental2023}.}

% \vspace{-2mm}
% \vspace{-4mm}
\section{Generative AI Usage Among SE Students (\textbf{RQ1})}
\label{RQ1_Results}
% In this section, we map the \textit{current usage of genAI amongst SE students}, where we identify how SE students use genAI to complement both conceptual learning and practical implementations. We present \textsc{when} and \textsc{how} students use genAI to complement their SE education, along with \textsc{where} they perceived benefits and challenges.

In this section, we map the \textit{current usage of genAI amongst SE students}, identifying \textsc{when} and \textsc{how} students use genAI to complement their SE learning and implementations, along with \textsc{where} they perceive benefits and challenges. 

\vspace{-1mm}
\subsection{\textbf{When do SE students use genAI?}} 
% Students use of genAI can be classified into two distinct phases: Learning and Implementation, where each phase is further subdivided into two.%:  Initial learning (L1) and Incremental learning (L2).

We explicitly inquired about students' use of genAI for learning and implementation phases. When analyzing their answers, we identified that each phase can be divided into two, as presented below. 
% The number in a grey box after the category name represents the number of participants who mentioned each theme---e.g., \numbox{10} means that 10 (out of 16) participants reported something about that.

%\boldification{Initial learning is when students start from scratch}

\textbf{Learning (Initial - L1)} corresponded to situations where participants used genAI to learn SE concepts from scratch, without any prior knowledge. Participants mentioned using genAI tools to get definitions and a basic understanding of SE concepts. For example, P9  \textit{``asked genAI to tell what fuzzing is and what its purpose was.''}

%\boldification{incremental is when students have some domain knowledge when using genAI}

\textbf{Learning (Incremental - L2)} is characterized by situations when participants, after having some background knowledge, used genAI to \textit{``refine understanding and tie up loose ends'' (P5)}. P2 mentioned that genAI was useful for \textit{``clarifying sprint and agile concepts that were confusing in the lecture'' (P2).} 

%\boldification{Basic implementations....}

\textbf{Implementation (Initial - I1)} involved participants' engagement with genAI during the initial stages of software implementations. They reported instances of using genAI to \textit{``whip up boilerplates'' (P15)}, set up frameworks, and assess code rundowns. P10 elaborated, \textit{``I used ChatGPT for setting up a quick react framework and got rundowns of structure and content of the code''}.

%\boldification{Advanced implementations...}
\textbf{Implementation (Advanced - I2)} corresponded to scenarios when participants used genAI to complement contributions to existing code bases or their own term/capstone projects. P1 used genAI for: \textit{``optimizing the solutions that I have built. I ask it for the recommended best practices or industry standard approach for the type of problem I am dealing with''}.

% Overall, participants indicated their use of genAI to complement conceptual learning in the initial (L1) and incremental phases (L2). For practical implementations, they engaged with genAI during the basic (I1) and advanced (I2) phases. 

\revised{Table \ref{tab:course-content} presents exemplary concepts and course contents corresponding to each of these phases, as provided by the course instructors. This can help readers relate our subsequent findings from an SE curriculum perspective.} For clarity and ease of reference throughout this paper, we will use the abbreviations L1, L2, I1, and I2 to refer to these phases.

\vspace{-4mm}
\begin{table}[!h]
\scriptsize
\centering
\caption{SE course contents corresponding to different phases of SE learning and implementations}
\vspace{-1mm}
\label{tab:course-content}
\begin{tabular}{p{0.4cm}p{3.6cm}p{3.6cm}}
\hline
\textbf{} & \textbf{Introductory/Intermediate} & \textbf{Senior/Capstone} \\ \hline \hline
\rowcolor{gray!20}\multirow{2}{*}{L1} & Version control systems, APIs and microservices, UI design principles, User stories, Scrum practices (e.g. sprints, backlogs), Code review 
& Project requirement analysis, Architectural patterns (MVC, MVVM), System design and scaling, SE best practices (code, documentation, review)  \\ \hline
\multirow{2}{*}{L2} & Advanced Git features (e.g. merging, rebasing, resolving conflicts), RESTful services, HTTP methods, Code smells, Software design, Software process models (principles, roles)
& Designing API gateways, Linters, Application performance monitoring, Software security, OAuth, Code performance optimization, System integration testing \\ \hline
\rowcolor{gray!20}\multirow{2}{*}{I1} & Setting up frameworks and environments, Dependency management, UI development (HTML, CSS, JS), Third party API integration, Unit tests 
& Software prototyping (design choices, core functionalities), Environment configuration for multiple service support, Setting automated test frameworks \\ \hline
\multirow{2}{*}{I2} & Building microservices, Testing frameworks,  Integration and end2end tests, Inter-process communication, Term project, Code reviews, PRs and issues 
& Software deployment (GH pages, AWS), Maintenance routines, CI/CD, Manage code conflicts across git branches, Term project \\ \hline
\end{tabular}
\vspace{-5mm}
\end{table}
% \vspace{-1.25mm}
\subsection{\textbf{How do they use it?}}

During the \textbf{learning} phase, participants described four ways in which they used genAI to support their learning.

\textit{1) \textbf{Role-based prompting}} involves instructing genAI systems to assume a specific role (e.g., a teaching assistant) or to explain concepts to a particular type of audience (e.g., a child). Participants mentioned that they prompted genAI tools to simplify complex concepts by requesting explanations in layman's terms or asking it to \textit{``explain things to a child'' (P8)}. Further,  participants gave tutoring roles to genAI to get deeper explanations on topics. For example, P2 shared, \textit{``I used to often ask ChatGPT and Gemini to pretend to be my TA for the class, pretending I'm a complete novice'' (P2)}. %to explain these things to me

\textit{2) \textbf{Using AI as an advanced search engine}} allowed participants to filter and \textit{``narrow down the search space'' (P12)}, pinpointing information for the topic of interest. For example, P9 mentioned, \textit{``I used to ask ChatGPT to get a basic idea of what all I needed to learn. If it gave me 3 things, I would then look up and read about those things''}. 
% Further, P12 mentioned that genAI tools aided in \textit{``narrowing down the search space, pointing to resources that best matched the concept I am trying to learn''}.

\textit{3) \textbf{Combining AI responses with other sources}} allowed for a more complete understanding. Participants considered genAI to be \textit{``not really good at giving the complete answer'' (P16)}. They mentioned combining AI responses with other sources such as Youtube (P16), Stack Overflow (P9), for a better understanding of SE concepts, instead of solely relying on genAI-provided explanations.

\textit{4) \textbf{Using course materials to query AI}} enabled participants to ask focused questions to refine or expand on their knowledge. For instance, P4 described, \textit{``I get familiarized with the jargon using class resources, and then come back to GPT and ask specific questions. I then use that knowledge to learn the concept and then apply it for my problem''}.

% For example, after roughly knowing what end to end tests are, I go to ChatGPT and prompt it saying this is my knowledge of end to end tests, can you please tell me more on [topic]?

% In discussing ways to complement software implementations with genAI tools, interviewees mentioned three distinct approaches:

Participants reported using genAI tools for supporting \textbf{implementations} in three distinct ways.

\textit{1) \textbf{Using AI to improve own or obtained code}}: Participants used genAI tools to refine their code (or the ones retrieved from other sources), for better performance and compliance with best practices. P1 explained, \textit{``When dealing with a problem, I use generative AI to optimize the solutions I have built. I ask it for the recommended best practices or industry-standard approach for this type of problem''}. 
%Similarly, P4 stated, \textit{``I go online to find sources with the code and ask ChatGPT to help me optimize that''}.

\textit{2) \textbf{Using AI for code snippets and organizing it on their own}}: Participants highlighted their use of genAI to acquire code components or \textit{``pieces of the puzzle''}, as P10 puts it, which they then organize into cohesive solutions for their projects. P3 shared, \textit{``I lean towards using genAI for individual parts that I planned dealing with in my project, that's where I think its value comes from...I can take those pieces and organize it to fit them into my project''.}

\textit{3) \textbf{Combining AI-generated code with traditional sources}}: Participants blended genAI-generated code with insights from traditional sources, like online documentation (P3), Google searches (P7), and forums such as Stack Overflow (P15), to craft effective solutions. This approach mirrors the \textit{combination of AI responses with other sources} for learning concepts.
% P3 described this approach, stating,  \textit{``I used ChatGPT for web scraping...and I used Mozilla developer docs to fulfill the rest if they exist''.} Similarly, 
P15 described this approach: \textit{``for me, the development cycle is to start with ChatGPT, refine it with documentation or specific Stack Overflow examples to fill the portions it missed''}.
% . They both have their purpose and are both useful in their own way''}.

\begin{figure*}[!bht]
\centering
% \vspace{-10px}
\includegraphics[width=0.85\textwidth]{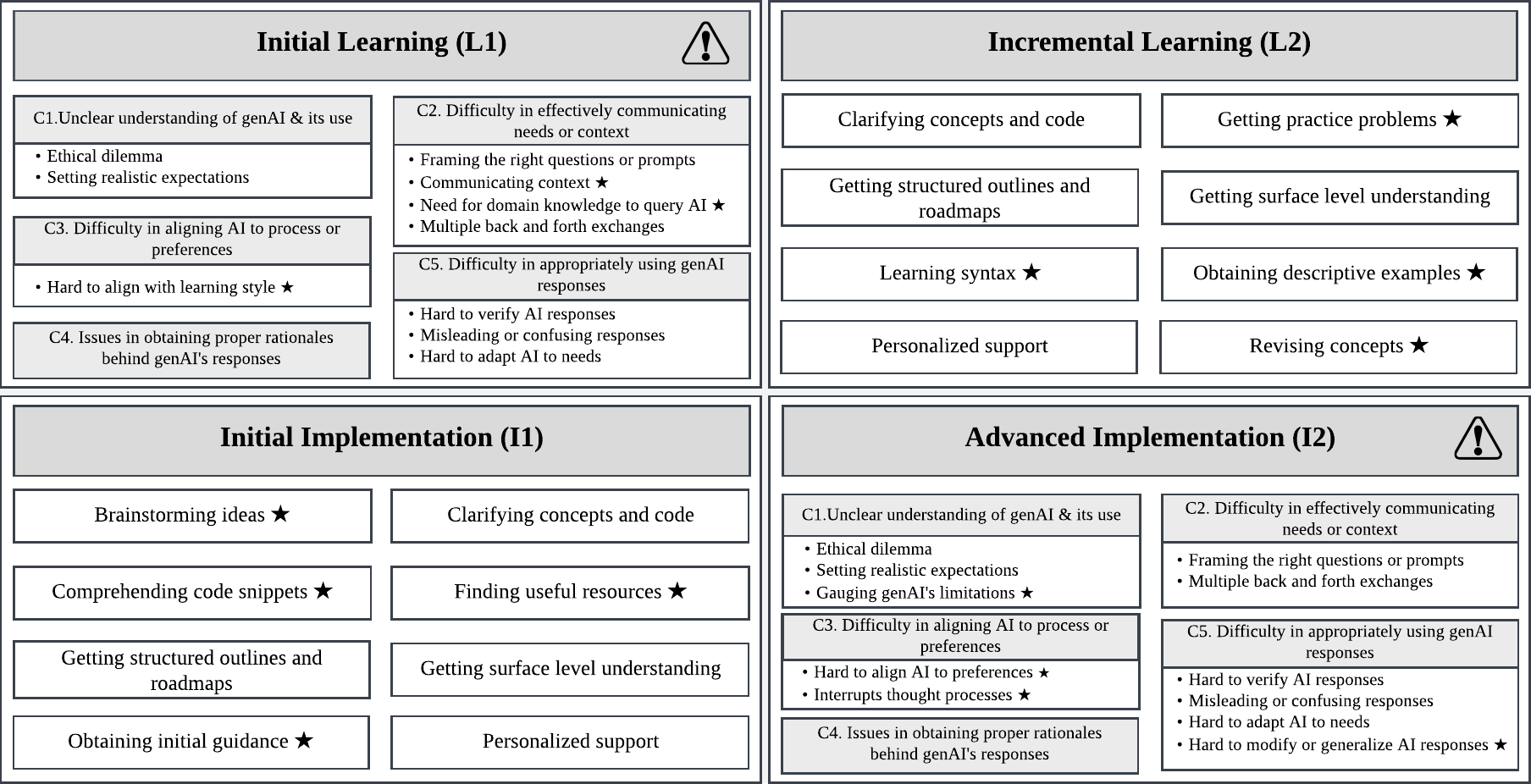}
\vspace{-8px}
\caption{\small {Perceived benefits and challenges of genAI usage among students: Benefits were in incremental learning (L2) \& initial implementations (I1), with challenges in initial learning (L1) \& advanced implementations (I2). Stars indicate perceptions unique to the quadrant.} }
\vspace{-8px}
\label{fig:landscape}
\vspace{-9px}
\end{figure*}

\vspace{-1mm}
\subsection{\textbf{Perceived benefits and challenges}}
\label{sec:benefits_challenges}

\vspace{-1mm}
% \boldification{Students perceive benefits, but these are only in L2 and I1 phases} 

Figure \ref{fig:landscape} presents an overview of the benefits and challenges perceived by SE students in using genAI tools, highlighting the specific phases in which these perceptions arise. Participants perceived the benefits of using genAI only for incremental learning (L2) and initial implementation tasks (I1). Conversely, they encountered challenges when using genAI for initial learning (L1) and advanced implementations (I2).

\textbf{\textsc{Benefits}}: Interviewees valued genAI tools both in incremental learning (L2) and initial implementations (I1) for its ability to provide \textit{personalized support} and \textit{clarify concepts and code}. One participant mentioned that \textit{``if I need a clearer explanation of a topic from class or if I am stuck with something in my code, I definitely turn to generative AI for a personalized explanation''} (P12). Moreover, genAI was valued for \textit{providing surface level understanding} and \textit{structured outlines or roadmaps} which aided in identifying key topics or areas of focus. P3 shared, \textit{``I find myself leaning more and more on genAI tools to get structured outlines and to know what information I need to gather''}.

In the incremental learning phase (L2), genAI proved beneficial for \textit{revising concepts}: \textit{``ChatGPT can help in brushing up or polishing concepts if you have a basic understanding of it'' (P14)} and \textit{obtaining descriptive examples}: \textit{``you can ask generative AI to give you an example or demonstrate what you're asking about in practice, which is helpful'' (P3)}. Participants also perceived these tools to be useful for \textit{getting practice problems} to enhance their understanding of a topic. P1 highlighted,  \textit{``I would also ask for incrementally challenging practice problems with a topic. It helped me have a practical understanding''.} Moreover, these tools were useful for \textit{learning syntax}, when participants had a programming background: \textit{``GenAI was useful when I needed to learn a particular JS syntax. I could understand what it said because I already knew how to program in JS'' (P13)}.

Interviewees also valued genAI tools during the initial implementations (I1) phase, particularly for \textit{comprehending code snippets}, \textit{brainstorming ideas}, and \textit{finding useful resources}. P4 stated that \textit{``given a problem or an assignment, the hardest part is to get started. AI is really helpful in that, it is useful in brainstorming, and it points you to sources that can lead you to the goal''}. Participants noted that they turned to genAI for \textit{initial guidance}, seeking instructions to follow and a code structure to work with. P9 shared, \textit{``I was clueless at the start of a game development project. I asked ChatGPT how to go through developing that and it gave me instructions on steps I needed to follow tailored to my project''}. 

% and a couple of steps that I needed to follow''}.

\revised{Instructors largely confirmed these benefits, with the exception of the value of ``personalized support'', where they emphasized that \textit{``AI doesn’t always get the context right or understand learning goals, making personalized [support] double-edged, it is risky when students build too much on flawed help'' (Capstone Instructor)}.}

\textbf{\textsc{Challenges}}: Interviewees highlighted various challenges in using genAI tools as complete beginners learning SE concepts (L1) and in tackling complex implementations (I2). These challenges can be organized into five main categories. First,
participants reported challenges in understanding genAI's limitations and identifying its appropriate and ethical applications \textbf{(C1)}. They found it difficult to effectively communicate their needs or context to the AI \textbf{(C2)} and to align it with their process or preferences \textbf{(C3)}. They also faced obstacles in obtaining clear rationales for the genAI responses \textbf{(C4)} and applying them appropriately \textbf{(C5)}. 

These categories were identified in both the L1 (initial learning) and I2 (advanced implementation) phases, with specific challenges occurring in either phase or spanning both. For clarity, we outline these distinctions in Figure \ref{fig:landscape} and proceed to explain the categories in the text:

\textit{C1) \textbf{Unclear understanding of genAI and its use}}: 
Participants reported difficulties understanding genAI's limitations and gauging its appropriate applications in software engineering. For example, they struggled to \textit{determine the ethical use} of genAI in SE: \textit{``I don't know whether it is ethical to use generative AI for actual implementations'' (P10)}, indicating moral ambiguity surrounding the tool's use. \textit{Setting realistic expectations} is another hurdle reported by our sample.  Participants' high expectations often clashed with genAI's actual capabilities, leading to dissatisfaction. P14's experience with genAI illustrates this gap: \textit{``I know for a fact that it wasn't exactly as helpful as I thought it would be...because it did not quite match my expectation''.} Lastly, \textit{gauging genAI's limitations} proved difficult for participants, particularly at the onset of use. P14 commented on their initial skepticism, \textit{``Initially, when I first started using it, I would trust it less because I wasn't quite exposed yet to how they work and how effective they actually were''}.

\textit{C2) \textbf{Difficulty in effectively communicating needs or context}}: Participants highlighted several challenges in effectively conveying their needs and contexts to genAI. \textit{Framing the right questions or prompts} emerged as an obstacle for those seeking useful responses. P6 highlighted, \textit{``My biggest challenge is asking it the right questions, to get a useful response out of it''.} 
Participants reported that \textit{communicating context} to genAI was challenging. P3 described the struggle: \textit{``The biggest challenge was providing enough context to [ChatGPT] and communicate my needs about what I'm trying to do''}. Furthermore, interviewees emphasized that achieving appropriate responses required engaging in iterative dialogues with genAI involving \textit{multiple back-and-forth exchanges}. P7 explained, \textit{``Sometimes, I just go back and forth with the context till it fixes its answers''}. Effective use of genAI also \textit{requires domain knowledge}, complicating its utility for those without it:  \textit{``The difficulty is when there is a knowledge gap and you are using it for things without having solid fundamentals about it. If you don't understand the concept, you can't use AI for it'' (P4)}.

\textit{C3) \textbf{Difficulty in aligning AI to process or preferences}}: Interviewees reported difficulties in \textit{aligning genAI to their preferences}. P16 articulated this issue, stating that genAI \textit{``does not take into account my preferences or style or what I want to fix in the codebase''}. Moreover, code-generating AI tools, such as Copilot, can \textit{interrupt thought processes} when it \textit{``gives huge blocks of things that you don't want, it breaks the flow and thought process'' (P7)}. Further, participants mentioned facing significant challenges in learning when genAI does not \textit{align with their learning style}. P12 reported that \textit{``generative AI was not aligned to my style of learning...the way it talks to me is not the way I want to learn''. }

\textit{C4) \textbf{Issues in obtaining proper rationales behind genAI responses}}: Interviewees noted their struggle to secure rationales for AI-suggested procedures and solutions. P1 stated, \textit{``If I want to know why a procedure operates the way it does, ChatGPT doesn't give me a full answer...when I am seeking a deeper understanding of what's going on''}. Participants shared that it was hard to discern errors with genAI and \textit{``to understand what was going wrong and why'' (P12)}. 

\textit{C5) \textbf{Difficulty in appropriately using genAI responses}}: Participants highlighted their struggles for appropriately using genAI responses in their work. \textit{Misleading or confusing responses} troubled them, as described by P5: \textit{``It outputted sensible looking code but it didn't match my instructions, leading to confusion''}. They found it hard to \textit{adapt genAI solutions to their context} or \textit{``make it generate content that fits all the needs''} (P3). Lastly, participants reported challenges in \textit{verifying genAI responses} alongside \textit{modifying or generalizing} them for wider use. P3 mentioned, \textit{``the responses were very hard to generalize or modify to include all cases''.}

\revised{Instructors validated these challenges without revealing any new ones. They also noted that students still remain hesitant to discuss challenges due to unclear academic AI policies.}

% They confirmed four sub-categories through personal experience and observations, noting that students remain hesitant to discuss challenges due to unclear AI policies.}

% Overall, while there are benefits, there are also challenges. The timing of introducing GenAI is crucial to maximizing its benefits.

\boldification{Overall, in the landscape, genAI is useful for L2 and I1 and not otherwise}

% Overall, generative AI is valued by SE students when they possess some domain knowledge (L2) or are engaged in early-stage implementations (I1). While it serves as a useful tool in certain contexts, its value diminishes for beginners (L1) and in complex tasks (I2).

%  ==========================End of RQ1======================
% \vspace{1mm}

\section{Antecedents \& Impacts of Challenges (\textbf{RQ2})} 
\label{RQ2_Results}

\begin{figure*}[!bht]
\centering
% \vspace{-10px}
% \includegraphics[width=0.85\textwidth]{figures/graph-edu.png}
\includegraphics[width=0.8\textwidth]{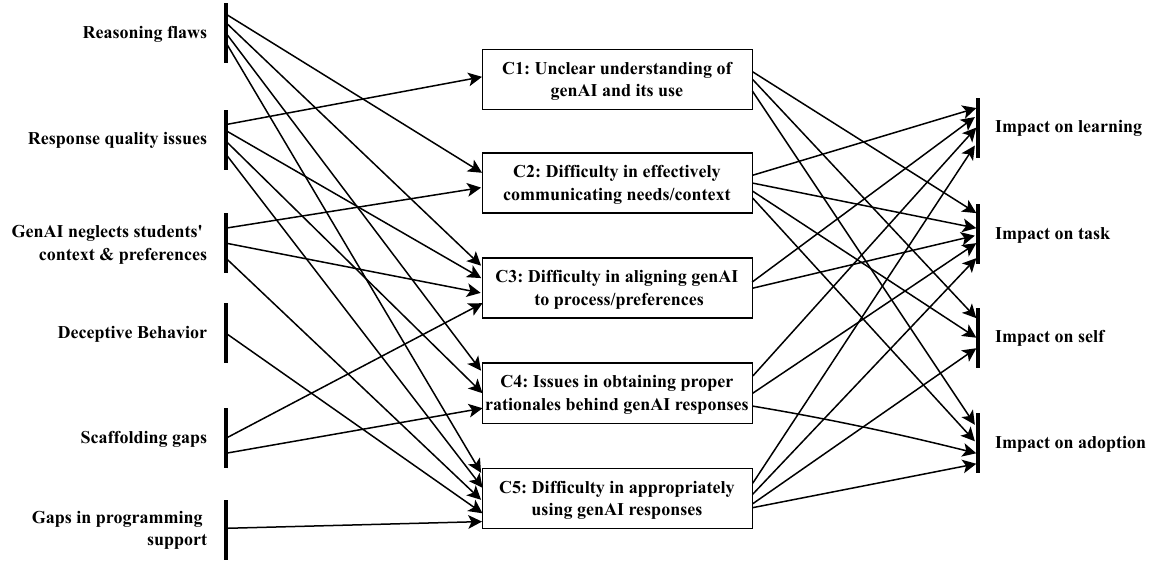}
\vspace{-8px}
\caption{\small {Associations between genAI's intrinsic issues (faults and gaps), challenges (C1-C5), and the resulting impacts. Issues in genAI contributed to various challenges encountered by participants, which subsequently had impacts on learning, task, self, and genAI adoption.} }
\vspace{-5px}
\label{fig:sankeys}
\vspace{-10px}
\end{figure*}

In this section, we answer RQ2, detailing the causes and consequences of the challenges encountered by SE students using genAI tools. Figure \ref{fig:sankeys} illustrates the associations between genAI intrinsic issues (causes), challenges, and subsequent impacts (consequences), which we detail next:

\boldification{Students identified issues with AI that they perceived contributed to their challenges. Here we explain the associations between AI deficiencies and human challenges}

\vspace{-1mm}
\subsection{\textbf{Causes}}  
Interviewees identified several intrinsic issues with genAI that contributed to their challenges, which we classified into faults and gaps. Faults refer to flaws within genAI systems, while gaps highlight areas where genAI is currently lacking.

\boldification{Participants highlighted several faults and gaps, we group faults and discuss their associations with human challenges}

\textsc{\textbf{Faults}}: Participants highlighted 13 different AI faults, which fall under four different categories:
% , discussed in the following:

\textit{1) \textbf{Reasoning flaws}} in genAI manifest notably through \textit{logic loops} and \textit{self-contradiction}. Logic loops occur when genAI is stuck in repetitive cycles of responses, failing to advance toward a solution that is appropriate. Self-contradiction involves genAI contradicting its statements or presenting \textit{``conflicting information'' (P4)} with its earlier conditions or premises. P7 illustrated a typical scenario: \textit{``sometimes AI will give you the wrong piece of code, and then you'll try it. It doesn't work...oh, I see the mistake and then will give you the same piece of code back...I just go back and forth with it and that doesn't lead me anywhere''}. These reasoning flaws created significant hurdles for participants in communicating their needs effectively to genAI \textbf{(C2)}. These issues also led to difficulties in aligning genAI to match personal preferences \textbf{(C3)}: \textit{``It ignores constraints and gives the same answer again and again'' (P16)}, obtaining clear rationale behind responses \textbf{(C4)} and appropriately using genAI outputs \textbf{(C5)}: \textit{``GPT kept going back and back...then I was confused about how to use what it said and couldn't understand why that worked'' (P12)}. 

\textit{2) \textbf{Response quality issues}} encompass several problems with genAI outputs, including \textit{incomplete assistance}, \textit{inconsistent responses}, \textit{incompatibility}, \textit{suboptimal responses}, \textit{limited help on specifics}, and \textit{non-generalizability} of solutions.
 % See leftover, if there is some definition needed for this category.   

%GenAI's incomplete assistance resulted in an unclear understanding of its limitations, posing 
GenAI's incomplete assistance resulted in challenges related to setting appropriate expectations \textbf{(C1)}. For instance, P14 mentioned, \textit{``ChatGPT was not as helpful as I expected it to be...I needed a precise explanation for each step and it wasn't providing me with that, as it would skip through certain steps''}. Moreover, genAI's limited help on specifics made it difficult for participants to obtain proper rationales \textbf{(C4)}. P1 shared, \textit{``if I wanted to know why a procedure operates the way it does, rather than what I would envision in my head to be a better way, generative AI didn't give me a full answer which was annoying as I wanted a deeper understanding of what was going on''}. Additionally, other issues like incompatibility and non-generalizability posed extra challenges in aligning AI outputs with personal processes \textbf{(C3)} and appropriately using AI responses \textbf{(C5)}, as P3 pointed out, \textit{``It was hard to build general purpose software with GPT...it provided code that was not really reusable''}.

\textit{3) \textbf{Deceptive behavior}} in genAI is characterized by its \textit{hallucination} and \textit{confirmation bias} tendencies. These issues differ from mere response quality problems---such as incomplete or incorrect information---due to genAI's active efforts to persuade or convince the user. GenAI often generates plausible yet entirely fabricated information, leading to situations where \textit{``it gives baseless answers with convincing explanations...I never end up learning from it because there's a debate between what the AI said and the original source, and I don't know why something is correct''} (P3). Further, genAI often affirms user beliefs regardless of their validity. P4 mentioned, \textit{``AI won't take any accountability and would just apologize and agree to whatever I am saying''}, highlighting difficulties in discerning reliable assistance from misleading confirmations \textbf{(C5)}.

\textit{4) \textbf{GenAI neglects students' context and preferences,}} which manifests in various forms. It often \textit{neglects preferences, misinterprets students' problem}, and enforces \textit{misguided guardrails}, which, ultimately, leads to responses that ignore unique contexts. This was what P5 reported: \textit{``It doesn't want to give me output because it thinks it is a terms of use issue without evaluating my context''}. This kind of faults lead to challenges in communicating needs \textbf{(C2)}, and aligning AI to preferences \textbf{(C3)}, as P16 illustrated, \textit{``when you ask ChatGPT or any other genAI to not do something a particular way, it does it anyways''}. Further, participants reported that these issues led to challenges in appropriately using AI responses \textbf{(C5)} in their work. P9 shared: \textit{``after I explained repeatedly that I needed to optimize my script without changing external dependencies, the solutions suggested by Gemini and ChatGPT still involved alternate libraries''}.% ... I couldn't use what it gave at all''}.

\textsc{\textbf{Gaps}}: Interviewees highlighted six different gaps (areas where genAI lacks offerings) categorized as follows:

\textit{1) \textbf{Scaffolding gaps}} refer to the current shortcomings of genAI tools (as of 2024) in supporting learning. Participants reported that these tools fell short in \textit{providing an in-depth understanding of concepts} leading to challenges in obtaining proper rationales \textbf{(C4)}: \textit{``It was in [SE2] class...GPT or Bard didn't provide a specific reason that I felt was good enough or believable on why [a test case] should be designed that way'' (P2)}. Some participants noted that these tools \textit{lack visualization support for explaining concepts} (in free versions (e.g., ChatGPT) as of 2024) 
and considered these to be \textit{``not very helpful for visual learners'' (P16)}. Moreover, participants highlighted that genAI's tendency of \textit{giving out direct answers} often conflicted with their learning preferences \textbf{(C3)}, \textit{``ChatGPT does take away from my learning, just because it gives me that immediate answer without me having to find it. I wait before I ask a super big question just so I can think about it before I see a direct answer''} (P12). 

\textit{2) \textbf{Gaps in programming support}}: Participants pointed out gaps in genAI's ability to assist with programming tasks, notably its \textit{inadequate debugging support}, \textit{gaps in orchestrating code}, and its \textit{unsuitability for complex programming}, often attributed to genAI's limited working memory or outdated training information \cite{biswas2023evaluating}. Participants noted that genAI frequently faltered in \textit{``debugging functions or programs in general''} (P10) and struggled to \textit{``orchestrate code from different functions, files, or services together''} (P7), complicating the effective use of its responses \textbf{(C5)}. For complex programming tasks, genAI's support falls short, as described by P4: \textit{``working with ChatGPT or using GPT generated code for bigger projects or large apps is hard, it can break the codebase''}.

\vspace{-2mm}
\subsection{\textbf{Consequences}} Interviewees highlighted several consequences as a result of facing challenges in using genAI within the SE context:

\boldification{Facing challenges impacted learning, tasks, tool adoption, and overall self. We discuss the consequences and its relation with challenges}

\textit{1) \textbf{Impact on learning}}: 
% stems from challenges in effectively communicating needs or context (C2), aligning genAI to preference or process (C3), obtaining proper rationales behind genAI responses (C4) and effectively using them (C5), leading to students learning wrong information or having partial/incomplete understanding of concepts. 
Challenges in effectively communicating needs or context \textbf{(C2)} resulted in participants \textit{learning wrong information}. P1 reflected on this, stating, \textit{``Sometimes, genAI gives wrong information due to mistakes in how I've inquired as I didn't know enough about the problem to ask the question properly. I had to unlearn what AI said and relearn it using YouTube''}. Still, participants reported that difficulties in aligning genAI with learning preferences \textbf{(C3)} and obtaining proper rationales \textbf{(C4)} led to an \textit{incomplete understanding of concepts}, hindering in-depth comprehension: \textit{``I am not able to use AI to wrestle with the ideas in the way I like to learn and understand the reasoning behind things. Using genAI is just like being told what to do, I don't like that''} (P1). The participants also reported that lack of clear explanations from genAI impedes grasping the underlying principles of concepts, as on P6 words: \textit{``It didn't really seem to help me understand the critical components of why something might work and a lot of things went over my head''}. Moreover, participants indicated difficulties in verifying AI responses \textbf{(C5)}, particularly as beginners (L1), leading to uncertainty about the correctness of the learned material. P8's experience underlined this: \textit{``I ended up learning the wrong things often in the past...so when it gives me responses that are not there on other platforms, I am skeptical about whether I learned the right thing.''}

\revised{These impacts on learning can be viewed through the lens of Cognitive Load Theory (CLT) \cite{plass2010cognitive}. Essentially, difficulties in communicating needs (C2), aligning AI with preferences (C3), issues in obtaining clear rationales (C4), and verifying AI responses (C5) all add to extraneous cognitive load, where students must expend additional effort to use genAI and interpret its responses, detracting from their learning process. Moreover, challenges in obtaining proper rationales (C4) make it harder to grasp underlying principles (or reasoning), thus hindering in-depth comprehension.}

\textit{2) \textbf{Impact on task}}: 
% We identified three primary impacts on participants' task outcomes upon using genAI: it often led to \textit{slowdowns} or caused them to \textit{figure things out on their own} or \textit{abandon tasks} entirely. 
Challenges in obtaining proper rationales \textbf{(C4)} and effectively using AI responses \textbf{(C5)} forced participants to \textit{figure things out on their own}, leading to time-consuming setbacks: \textit{``GPT struggled to give an answer that worked or explain why it didn't work, it kept going back and back ... I ended up doing it again from scratch using a different resource'' (P12)}. Difficulty in aligning genAI to processes or preferences \textbf{(C3)} not only \textit{slowed them down} but also led to \textit{abandonment of tasks} in certain situations (I2). P16 shared, \textit{``ChatGPT didn't really give me a good answer. It went on for a long period of time, suggesting things that were similar but not useful to my project. I gave up and had to downsize that project''}, illustrating how genAI derailed their project goals and affected their timeline. Effectively communicating needs \textbf{(C2)} affected participants who spent substantial time refining prompts for actionable advice. P8 said that they \textit{``spent at least 30 minutes fixing prompts and going back and forth to make it work''}. Lastly, an unclear understanding of genAI's limitations \textbf{(C1)} often led students, particularly the beginners (L1), to give up on their task, when they found obstacles, due to lack of clear expectations about genAI's applicability. P12 clearly walked us through that: \textit{``For someone new, AI won't work, it would cause roadblocks and they would give up lacking an assumption of what it can be used for and what not.''} 
\revised{Extraneous load from ineffective AI responses (e.g., P12's need to redo tasks from scratch due to unsatisfactory explanations) and continuous prompt refinement (e.g., P8's time spent refining prompts) distract students from focusing on the task at hand. Further, task abandonment by P12 and P16 may be attributed to the combined intrinsic load (inherent difficulty of the task) and extraneous load (setting proper expectations about genAI's applicability and managing its responses), both of which impeded task completion.}

\textit{3) \textbf{Impact on self}}: Participants shared how they were personally impacted by using genAI, highlighting instances of \textit{over-reliance, self-doubt}, and \textit{frustration}. Over-reliance is common among beginners (L1) \cite{chen2021evaluating} due to a lack of understanding of genAI's limitations or its use \textbf{(C1)}, leading to over-dependence and initial blind trust in genAI. This was pointed out by P7, who mentioned that \textit{``Initially, I would just prompt it and trust whatever came back...I still find it confusing to realize whether GPT can actually provide a solution or just is claiming to do so''}. P15 added: \textit{``I think that new people should kind of stay away from AI as it is tempting to trust it blindly. If this black box is infallible to you, what happens when it's wrong?''}. Challenges in effectively communicating needs \textbf{(C2)} and using AI appropriately \textbf{(C5)} contributed to self-doubt and frustration, as expressed by P15, \textit{``It wasn't helpful because it made false assumptions about the codebase...the provided code was profoundly unhelpful, I couldn't modify it. It was very frustrating and just hopeless''}. %Still, P3 highlighted the struggles of using genAI for generating content \textit{``even after spending time learning prompt engineering''}. 
\revised{These experiences align with the concepts highlighted in the Social Cognitive Theory (SCT) \cite{bandura1986social}, where repeated failures erode users’ confidence and self-efficacy. When students repeatedly experience errors or incorrect outputs from genAI, it contributes to a cycle of frustration and self-doubt.} Finally, P4's comparison with peers underlines the impact on their self-esteem: \textit{``my friends can use it well, but I can't...I am worried I would be left behind when it replaces Google search and I don't know how to talk to it.''} \revised{This aligns with Self-Determination Theory (SDT) \cite{deci2012self}, which emphasizes the need for competence and relatedness. When users feel less competent than their peers, their intrinsic motivation and self-esteem suffer, leading to reduced engagement and satisfaction}.

\textit{4) \textbf{Impact on genAI adoption}}: Participants' willingness to adopt genAI was mainly influenced by challenges in appropriately using its responses \textbf{(C5)}. Yet, specific challenges impacting their willingness varied depending on the context of use. For learners (L1), difficulties in effectively communicating needs \textbf{(C2)} and obtaining proper rationales \textbf{(C4)} led to a sense of \textit{distrust}. P2 noted, \textit{``I don't trust generative AI for any critical work, it doesn't provide proper justifications; it's more like a predictive text answer that is hard to verify''}. P7 added: \textit{``It apologizes when you confront it, but gives you the same answer again and again. It's better to trust yourself and not the tool''}. These experiences revealed a broader skepticism towards genAI, deterring its adoption among SE students.
Further, participants' concerns over ethical considerations \textbf{(C1)} \textit{steered them away from using genAI} in advanced implementations (I2). P10 articulated this concern: \textit{``I don't use it for any implementations because I don't consider it to be something I have written...and I think until the boundaries become clearer on whether it's right to use it or not, I am not comfortable using it for work that I have to claim as my own''}. 

\revised{The impact of these challenges on genAI adoption can be understood through the Technology Acceptance Model (TAM) \cite{davis1989technology}, primarily through two constructs: perceived ease of use (PEOU) and perceived usefulness (PU). Both PEOU and PU influence users' intentions to use software tools. In this context, challenges in using genAI responses (C5) diminish PEOU, particularly for learners (L1), who struggle with communicating needs (C2) and receiving proper rationales (C4). Moreover, ethical considerations (C1) negatively influence both PU and PEOU, steering students away from adopting genAI for advanced implementations (I2).}

\revised{As with RQ1, we validated these findings with instructors, who confirmed the patterns without proposing any changes. The codebook is available in the supplemental material \cite{supplemental2023}.}

%  ==========================End of RQ2======================

% \vspace{-2mm}
\vspace{-5mm}
\section{Discussion}
\label{sec:discussion}
We identified different ways SE students make use of genAI, alongside where they perceive benefits and hurdles when using it to complement SE education. It is important to carefully consider how to incorporate genAI into the curriculum and its impact on students; as an instructor reflected:

\textit{``Maybe I thought of my students to be confident in figuring out the AI tools, I didn’t predict they would internalize these struggles and get frustrated...It’s more serious than I predicted and can be a very stressful experience for them."}

Backed by our results, we present a set of recommendations for integrating genAI in the SE education context. 

\noindent\textbf{\textit{Considerations before adopting genAI in SE education.}} 
Educators aiming to incorporate genAI into their SE curricula should be aware of the issues that students might encounter when using these tools for initial learning (L1) and advanced implementations (I2), as depicted in Figure~\ref{fig:landscape} and detailed in Section~\ref{sec:benefits_challenges}. A significant obstacle in adopting this rapidly developing technology is understanding and creating awareness of its limitations and applicability (C1). This can be particularly problematic for novice learners, who might rely too heavily on genAI, viewing its outputs as final or unquestionable truths \cite{chen2021evaluating}. Educators must, thus, educate themselves and the students about the workings and faults of genAI, specifically that these tools are language models and not knowledge models (i.e., they generate a sequence of symbols and not necessarily factual information). Several universities have put forth resources to help instructors understand how to manage/identify genAI usage in the classroom \cite{Harvard_OUE_2023, UCLA_CAT_2023}, but these largely explain how to interact with the tool (C2) and not how to incorporate its responses (C5). Educators must emphasize the need to verify genAI's outputs. This will help especially those starting in the course by (1) setting clear expectations regarding genAI's assistance, and (2) scoping their trust in AI \cite{amoozadeh2024trust, choudhuri2024guides}, instead of treating it like an expert.

Further, instead of penalizing students for using genAI, which would only lead to more creative ways of academic misconduct \cite{denny2024computing}, what is needed is to educate them on ethical and appropriate genAI usage and encourage them to acknowledge the extent of AI's role in their work. Students should also recognize that they are ultimately responsible for the code they generate---with (or without) genAI help. Thus, if security flaws or faults exist in the code, they need to take ownership. This ethos is particularly important to instill in senior SE students as they work with external clients (capstone projects), open source projects, or internships.

\noindent\textbf{\textit{Equip students with the right skills.}} Educators must focus on equipping SE students to be savvy-AI users, given the software industry's growing interest in integrating genAI into its practices \cite{bird2022taking}. While AI will not (potentially) supplant software engineers, individuals proficient in these technologies will have more job opportunities and career advancement. This leads to two guiding questions:

% 
%\footnote{Prompt engineering \cite{white2023prompt} involves structuring queries in a way that helps genAI better understand the context, leading to more relevant outputs.} 
% 
\textit{How to teach SE students to be savvy-AI users?} 
Prior research has frequently discussed the significance of prompt engineering \cite{white2023prompt} in leveraging genAI. Although it helps (instances being role-based prompting), it alone can't completely solve the problem. One of our participants (P3) highlighted struggles with prompt engineering: \textit{``even after spending time learning prompt engineering I couldn't make it listen to me''}.
To become savvy-AI users, students need to be taught the nuances of interacting with genAI: how to create different types of queries, how to effectively articulate the problem context to the AI, and that an effective prompt might require fundamental knowledge of SE concepts. Additionally, students need to be taught how to perform critical thinking to evaluate genAI outputs. They should be taught how to iteratively interact with the tool asking it further details or specific questions, and how to triangulate the AI responses with verifiable sources.  

\textit{How to adapt SE courses to genAI?} In hindsight, we now know the concerns in the 1940's about calculators undermining math education were not a reality. Instead, calculators became a useful tool in solving higher-order math problems. Today, genAI tools raise similar yet serious concerns, involving education and practices that require problem-solving 
% The use of genAI can reduce problem solving and ideation in creating code solutions into mere reading (code) comprehension exercise 
\cite{ko_2024}. This is problematic in SE education, which involves understanding the problem space, trade-offs, and contextual decision-making. So, \textit{what can we do about this?}
 First, just like students in primary school still need to learn to do basic math without calculators, complete novices need to be encouraged to learn SE topics without the aid of genAI tools. Second, educators must scaffold the introduction of genAI usage in their curricula, redesigning their assignments that place higher emphasis on reflection and problem-solving skills.  
%while revising the curricula to recognize the role of AI support, placing an emphasis on students' reflective skills.
For example, in addition to submitting code, students can be asked to reflect on their design decisions/program rationale/alternate choices that are connected back to the lectures. Asking for explanations for why the AI tool gave a particular solution and why that solution is a right choice (or not)---an area where genAI is ``stubbornly opaque'' \cite{amoozadeh2024trust}---will shift the focus from ``obtaining just a solution'' to critical thinking about the solution. Such reflections on alternate solutions are skills that SE students need to learn to be ready for an AI-infused workplace.
\vspace{-1mm}
\section{Threats to Validity}
\label{sec:threat}
% \vspace{-1mm}
% In this section, we report the limitations of our work.%\cite{wohlin2012experimentation}.

% When dealing with interviews, one important threat regards the accuracy of constructs used in the script, which may result in asking incorrect or ambiguous questions.
One important threat in interview studies is obtaining accurate constructs through questions and avoid asking ambiguous ones. 
% One important threat in interview studies regards the accuracy of constructs used in the script, to avoid asking incorrect or ambiguous questions.
To minimize this threat, we thoroughly evaluated our script internally and assessed its validity through sandboxing and pilot sessions until a team consensus was reached. 

Another threat could arise from the absence of concrete tasks conducted by the students during the study, limiting our results to our participants' recollections. We reduced this by focusing discussions on concrete artifacts to elicit participant experiences. Despite these measures, we recognize the potential for measurement inaccuracies.

%\textbf{\textit{Internal Validity}}: 
It is difficult to replicate qualitative research since human behaviors, feelings, and perceptions change over time. Thus, in the reliability thread, ~\citet{merriam2015qualitative} suggest checking the consistency of the inferences. To increase consistency, the analysis was conducted by two authors, followed by meetings with the team to refine and discuss the themes and categories until we reached a consensus. We also performed member checking and triangulated with instructors to validate the authenticity of our data interpretations and findings. 

Our study, like others, was susceptible to self-selection bias, where individuals interested in genAI might be more inclined to participate. Desirability bias might also have impacted our study, where participants might favor genAI due to its current popularity. To mitigate this, we framed our questions in neutral, unbiased language, encouraging participants to draw from actual experiences by providing concrete examples. Still, we analyzed the data cautiously acknowledging the potential influence of desirability bias while interpreting our results. Finally, despite our explicit efforts to assure participants about the confidentiality of the study, it is plausible that concerns over possible repercussions regarding genAI's academic use might have deterred them from fully disclosing their thoughts. 

% All participants interacted with the genAI tools in English. We didn’t observe any challenges due to language barriers in our sample. 

%\textbf{\textit{External Validity}}: 
Our study included participants from various courses and settings (in-person and online classes), still it is limited to a single university. Additionally, all participants interacted with the genAI tools in English and faced no language barriers.
Our findings, thus, may not generalize to other contexts or domains. 
% Our study investigates academic experiences of genAI usage among SE students. Although participants were recruited from across various courses and settings (in-person and online classes), our study is limited to a single university. 
% We traded off generalizability to mitigate biases associated with sampling due to inconsistencies in academic practices across different institutions. 
% Therefore, caution is advised when generalizing our findings to other contexts or domains. 
% Given this confinement, caution is advised when extrapolating our findings to broader contexts or domains. 
%A relatively small sample size of 16 participants may pose a potential threat. However, the quality, rather than the size, is essential to increase our confidence in the findings. Though we reached saturation after the $10^{th}$ interview, we proceeded with six additional interviews to ensure a diverse and balanced representation of perspectives. Still, we can not claim our findings will generalize across all settings. 
% 
Lastly, we report qualitative findings aggregated in emergent categories. While quantitative insights provide useful information, prior work has cautioned against using quantitative research methods on qualitative data \cite{denzin2011sage}. Therefore, we eschewed from discussing frequency or percentage of occurrences of categories. 

% \vspace{-3mm}
\vspace{-1.5mm}
\section{Conclusion}
\label{sec:conclusion}

Our study explored the current state of genAI usage among SE students to understand how they complement their SE learning and implementation tasks. Through 16 reflective interviews, we uncovered the contexts where these tools are helpful and where they pose challenges, examining why these challenges arise and how they impact students. Participants reported that they perceived benefits when using genAI for incremental learning (L2) or initial implementation tasks (I1). Conversely, challenges arose in using these tools as beginners (L1) or in advanced implementations (I2). Several challenges occurred due to intrinsic issues in genAI itself, leading to impacts on learning, tasks, self-perception, and genAI adoption. 

%\revised{\textit{``It's like learning to play the piano---once you hit the wrong key, it becomes hard to stop because what you've learned at that moment is what you learn the best.'' (SE-Intermediate Instructor)}}

A key call to action is on how to instill critical thinking, problem-solving abilities, and AI literacy among SE students both within traditional academic pathways and those aspiring to enter the field of SE.
% such that students' genAI learning is integrated with the software engineering knowledge.
Without clear guidelines on how to effectively use genAI, we may have a scenario akin to \textit{``learning to play the piano---once you hit the wrong key, it becomes hard to stop, as what you’ve learned in that moment is what you learn best.'' (SE-Intermediate Instructor)}

%Our findings provide a mapping for SE educators on where genAI can be incorporated along with its benefits and pitfalls. 

% OR

% Our study explored the current state of genAI use among SE students for SE learning and implementations, identifying benefits in incremental learning and initial implementations, but challenges in initial learning and advanced implementations. We found intrinsic genAI issues led to challenges, that further affected learning and task outcomes, self-perception, and willingness to adopt genAI. 

% Our findings urge SE educators to carefully integrate genAI into curricula, guiding ethical use of AI among students and adapting courseware to cultivate critical thinking, problem-solving, and AI literacy. This can prepare students—both within traditional pathways and those aspiring to enter the field of SE—become AI-savvy software engineers.

% to prepare students for AI-informed software engineering.

\noindent\textbf{Data Availability}: The research artifacts for this study are publicly available on the companion website \cite{supplemental2023}.

\noindent\textbf{Acknowledgements}: We thank all the study participants. This work was partially supported by the National Science Foundation under Grant Numbers 2235601, 2236198, 2247929, 2303042, and 2303043. %Any opinions, findings, conclusions, or recommendations expressed are those of the authors and do not necessarily reflect the views of the sponsors.

%%
%% The acknowledgments section is defined using the "acks" environment
%% (and NOT an unnumbered section). This ensures the proper
%% identification of the section in the article metadata, and the
%% consistent spelling of the heading.

% \begin{acks}
% We thank Samarendra Hedaoo for his insights in the meetings about the SE course. We thank all participants who took part in the study for their time and effort. 
% This work was partially supported by the National Science Foundation under Grant Numbers: 2235601, 2236198, 2247929, 2303042, and 2303043. Any opinions, findings, conclusions, or recommendations expressed in this material are those of the authors and do not necessarily reflect the views of the sponsors.

% We thank the GitHub Next team, Tom Zimmermann, and Christian Bird for providing valuable feedback on the survey contents and Brian Houck for facilitating the survey distribution. We also thank all the survey respondents for their time and insights. This work was partially supported by the National Science Foundation under Grant Numbers 2235601, 2236198, 2247929, 2303042, and 2303043. Any opinions, findings, conclusions, or recommendations expressed in this material are those of the authors and do not necessarily reflect the views of the sponsors.
% \end{acks}

% \bibliographystyle{ACM-Reference-Format}

\bibliographystyle{IEEEtranN}
% \bibliography{REFERENCES}
\footnotesize{\bibliography{acmart}}

% Generated by IEEEtranN.bst, version: 1.14 (2015/08/26)
\begin{thebibliography}{80}
\providecommand{\natexlab}[1]{#1}
\providecommand{\url}[1]{#1}
\csname url@samestyle\endcsname
\providecommand{\newblock}{\relax}
\providecommand{\bibinfo}[2]{#2}
\providecommand{\BIBentrySTDinterwordspacing}{\spaceskip=0pt\relax}
\providecommand{\BIBentryALTinterwordstretchfactor}{4}
\providecommand{\BIBentryALTinterwordspacing}{\spaceskip=\fontdimen2\font plus
\BIBentryALTinterwordstretchfactor\fontdimen3\font minus \fontdimen4\font\relax}
\providecommand{\BIBforeignlanguage}[2]{{%
\expandafter\ifx\csname l@#1\endcsname\relax
\typeout{** WARNING: IEEEtranN.bst: No hyphenation pattern has been}%
\typeout{** loaded for the language `#1'. Using the pattern for}%
\typeout{** the default language instead.}%
\else
\language=\csname l@#1\endcsname
\fi
#2}}
\providecommand{\BIBdecl}{\relax}
\BIBdecl

\bibitem[OpenAI(2024)]{GPT4}
OpenAI, ``Gpt-4,'' \url{https://openai.com/product/gpt-4}, 2024.

\bibitem[Google(2024)]{Gemini}
Google, ``Gemini,'' \url{https://gemini.google.com}, 2024.

\bibitem[Microsoft(2024)]{Copilot}
Microsoft, ``Copilot,'' \url{https://copilot.microsoft.com}, 2024.

\bibitem[Fan et~al.(2023)Fan, Gokkaya, Harman, Lyubarskiy, Sengupta, Yoo, and Zhang]{fan2023large}
A.~Fan, B.~Gokkaya, M.~Harman, M.~Lyubarskiy, S.~Sengupta, S.~Yoo, and J.~M. Zhang, ``Large language models for software engineering: Survey and open problems,'' in \emph{2023 IEEE/ACM International Conference on Software Engineering: Future of Software Engineering (ICSE-FoSE)}.\hskip 1em plus 0.5em minus 0.4em\relax IEEE, 2023, pp. 31--53.

\bibitem[Ebert and Louridas(2023)]{ebert2023generative}
C.~Ebert and P.~Louridas, ``Generative {AI} for software practitioners,'' \emph{IEEE Software}, vol.~40, no.~4, pp. 30--38, 2023.

\bibitem[Russo(2024)]{russo2023navigating}
D.~Russo, ``Navigating the complexity of generative {AI} adoption in software engineering,'' \emph{ACM Transactions on Software Engineering and Methodology}, 2024.

\bibitem[Bull and Kharrufa(2023)]{bull2023generative}
C.~Bull and A.~Kharrufa, ``Generative {AI} assistants in software development education: A vision for integrating generative {AI} into educational practice, not instinctively defending against it.'' \emph{IEEE Software}, 2023.

\bibitem[Nguyen-Duc et~al.(2023)Nguyen-Duc, Cabrero-Daniel, Przybylek, Arora, Khanna, Herda, Rafiq, Melegati, Guerra, Kemell, et~al.]{nguyen2023generative}
A.~Nguyen-Duc, B.~Cabrero-Daniel, A.~Przybylek, C.~Arora, D.~Khanna, T.~Herda, U.~Rafiq, J.~Melegati, E.~Guerra, K.-K. Kemell \emph{et~al.}, ``Generative artificial intelligence for software engineering--a research agenda,'' \emph{arXiv preprint arXiv:2310.18648}, 2023.

\bibitem[Denny et~al.(2024)Denny, Prather, Becker, Finnie-Ansley, Hellas, Leinonen, Luxton-Reilly, Reeves, Santos, and Sarsa]{denny2024computing}
P.~Denny, J.~Prather, B.~A. Becker, J.~Finnie-Ansley, A.~Hellas, J.~Leinonen, A.~Luxton-Reilly, B.~N. Reeves, E.~A. Santos, and S.~Sarsa, ``Computing education in the era of generative {AI},'' \emph{Communications of the ACM}, vol.~67, no.~2, pp. 56--67, 2024.

\bibitem[Welsh(2022)]{welsh2022end}
M.~Welsh, ``The end of programming,'' \emph{Communications of the ACM}, vol.~66, no.~1, pp. 34--35, 2022.

\bibitem[Yellin(2023)]{yellin2023premature}
D.~M. Yellin, ``The premature obituary of programming,'' \emph{Communications of the ACM}, vol.~66, no.~2, pp. 41--44, 2023.

\bibitem[Chen et~al.(2021)Chen, Tworek, Jun, Yuan, Pinto, Kaplan, Edwards, Burda, Joseph, Brockman, et~al.]{chen2021evaluating}
M.~Chen, J.~Tworek, H.~Jun, Q.~Yuan, H.~P. d.~O. Pinto, J.~Kaplan, H.~Edwards, Y.~Burda, N.~Joseph, G.~Brockman \emph{et~al.}, ``Evaluating large language models trained on code,'' \emph{arXiv preprint arXiv:2107.03374}, 2021.

\bibitem[Bommasani et~al.(2021)Bommasani, Hudson, Adeli, Altman, Arora, von Arx, Bernstein, Bohg, Bosselut, Brunskill, et~al.]{bommasani2021opportunities}
R.~Bommasani, D.~A. Hudson, E.~Adeli, R.~Altman, S.~Arora, S.~von Arx, M.~S. Bernstein, J.~Bohg, A.~Bosselut, E.~Brunskill \emph{et~al.}, ``On the opportunities and risks of foundation models,'' \emph{arXiv preprint arXiv:2108.07258}, 2021.

\bibitem[Amoozadeh et~al.(2023)Amoozadeh, Daniels, Chen, Nam, Kumar, Hilton, Alipour, and Ragavan]{amoozadeh2023towards}
M.~Amoozadeh, D.~Daniels, S.~Chen, D.~Nam, A.~Kumar, M.~Hilton, M.~A. Alipour, and S.~S. Ragavan, ``Towards characterizing trust in generative artificial intelligence among students,'' in \emph{2023 ACM Conference on International Computing Education Research-Volume 2}, 2023, pp. 3--4.

\bibitem[Becker et~al.(2023)Becker, Denny, Finnie-Ansley, Luxton-Reilly, Prather, and Santos]{becker2023programming}
B.~A. Becker, P.~Denny, J.~Finnie-Ansley, A.~Luxton-Reilly, J.~Prather, and E.~A. Santos, ``Programming is hard-or at least it used to be: Educational opportunities and challenges of {AI} code generation,'' in \emph{Proceedings of the 54th ACM Technical Symposium on Computer Science Education V. 1}, 2023, pp. 500--506.

\bibitem[Penney et~al.(2024)Penney, Pimentel, Steinmacher, and Gerosa]{penney2023conversations}
J.~Penney, J.~F. Pimentel, I.~Steinmacher, and M.~A. Gerosa, ``Anticipating user needs: Insights from design fiction on conversational agents for computational thinking,'' in \emph{Chatbot Research and Design}, A.~F{\o}lstad, T.~Araujo, S.~Papadopoulos, E.~L.-C. Law, E.~Luger, M.~Goodwin, S.~Hobert, and P.~B. Brandtzaeg, Eds.\hskip 1em plus 0.5em minus 0.4em\relax Cham: Springer Nature Switzerland, 2024, pp. 204--219.

\bibitem[Vaswani et~al.(2017)Vaswani, Shazeer, Parmar, Uszkoreit, Jones, Gomez, Kaiser, and Polosukhin]{vaswani2017attention}
A.~Vaswani, N.~Shazeer, N.~Parmar, J.~Uszkoreit, L.~Jones, A.~N. Gomez, {\L}.~Kaiser, and I.~Polosukhin, ``Attention is all you need,'' \emph{Advances in neural information processing systems}, vol.~30, 2017.

\bibitem[Achiam et~al.(2023)Achiam, Adler, Agarwal, Ahmad, Akkaya, Aleman, Almeida, Altenschmidt, Altman, Anadkat, et~al.]{achiam2023gpt}
J.~Achiam, S.~Adler, S.~Agarwal, L.~Ahmad, I.~Akkaya, F.~L. Aleman, D.~Almeida, J.~Altenschmidt, S.~Altman, S.~Anadkat \emph{et~al.}, ``Gpt-4 technical report,'' \emph{arXiv preprint arXiv:2303.08774}, 2023.

\bibitem[Choudhuri et~al.(2024{\natexlab{a}})Choudhuri, Liu, Steinmacher, Gerosa, and Sarma]{choudhuri2024far}
R.~Choudhuri, D.~Liu, I.~Steinmacher, M.~Gerosa, and A.~Sarma, ``{How Far Are We? The Triumphs and Trials of Generative AI in Learning Software Engineering},'' in \emph{Proceedings of the IEEE/ACM 46th International Conference on Software Engineering}, 2024, pp. 1--13.

\bibitem[Ko(2024)]{ko_2024}
A.~J. Ko, ``More than calculators: Why large language models threaten public education,'' Jan 2024.

\bibitem[Lau and Guo(2023)]{lau2023ban}
S.~Lau and P.~Guo, ``From" ban it till we understand it" to" resistance is futile": How university programming instructors plan to adapt as more students use {AI} code generation and explanation tools such as chatgpt and github copilot,'' in \emph{2023 ACM Conference on International Computing Education Research-Volume 1}, 2023, pp. 106--121.

\bibitem[Liang et~al.(2024)Liang, Yang, and Myers]{liang2023understanding}
J.~T. Liang, C.~Yang, and B.~A. Myers, ``A large-scale survey on the usability of {AI} programming assistants: Successes and challenges,'' in \emph{Proceedings of the 46th IEEE/ACM International Conference on Software Engineering}, 2024, pp. 1--13.

\bibitem[Whiting(2023)]{whiting_2023}
\BIBentryALTinterwordspacing
K.~Whiting, ``How {AI} is helping to identify skills gaps and future jobs,'' May 2023. [Online]. Available: \url{https://www.weforum.org/agenda/2023/05/ai-skills-gaps-future-jobs/}
\BIBentrySTDinterwordspacing

\bibitem[Woodruff et~al.(2024)Woodruff, Shelby, Kelley, Rousso-Schindler, Smith-Loud, and Wilcox]{woodruff2023knowledge}
A.~Woodruff, R.~Shelby, P.~G. Kelley, S.~Rousso-Schindler, J.~Smith-Loud, and L.~Wilcox, ``How knowledge workers think generative {AI} will (not) transform their industries,'' in \emph{Proceedings of the CHI Conference on Human Factors in Computing Systems}, 2024, pp. 1--26.

\bibitem[Prather et~al.(2023{\natexlab{a}})Prather, Denny, Leinonen, Becker, Albluwi, Craig, Keuning, Kiesler, Kohn, Luxton-Reilly, et~al.]{prather2023robots}
J.~Prather, P.~Denny, J.~Leinonen, B.~A. Becker, I.~Albluwi, M.~Craig, H.~Keuning, N.~Kiesler, T.~Kohn, A.~Luxton-Reilly \emph{et~al.}, ``The robots are here: Navigating the generative {AI} revolution in computing education,'' in \emph{Proceedings of the 2023 Working Group Reports on Innovation and Technology in Computer Science Education}, 2023, pp. 108--159.

\bibitem[Malik et~al.(2023)Malik, Dwivedi, Kshetri, Hughes, Slade, Jeyaraj, Kar, Baabdullah, Koohang, Raghavan, et~al.]{malik2023so}
T.~Malik, Y.~Dwivedi, N.~Kshetri, L.~Hughes, E.~L. Slade, A.~Jeyaraj, A.~K. Kar, A.~M. Baabdullah, A.~Koohang, V.~Raghavan \emph{et~al.}, ``“so what if chatgpt wrote it?” multidisciplinary perspectives on opportunities, challenges and implications of generative conversational ai for research, practice and policy,'' \emph{International Journal of Information Management}, vol.~71, p. 102642, 2023.

\bibitem[Rajabi et~al.(2023)Rajabi, Taghipour, Cukierman, and Doleck]{rajabi2023exploring}
P.~Rajabi, P.~Taghipour, D.~Cukierman, and T.~Doleck, ``Exploring chatgpt’s impact on post-secondary education: A qualitative study,'' in \emph{Proceedings of the 25th Western Canadian Conference on Computing Education}, 2023, pp. 1--6.

\bibitem[Joshi et~al.(2023)Joshi, Budhiraja, Tanna, Jain, Deshpande, Srivastava, Rallapalli, Akolekar, Challa, and Kumar]{joshi2023let}
I.~Joshi, R.~Budhiraja, P.~D. Tanna, L.~Jain, M.~Deshpande, A.~Srivastava, S.~Rallapalli, H.~D. Akolekar, J.~S. Challa, and D.~Kumar, ``From" let's google" to" let's chatgpt": Student and instructor perspectives on the influence of llms on undergraduate engineering education,'' \emph{arXiv preprint arXiv:2309.10694}, 2023.

\bibitem[Raman et~al.(2023)Raman, Mandal, Das, Kaur, Sanjanasri, and Nedungadi]{raman2023university}
R.~Raman, S.~Mandal, P.~Das, T.~Kaur, J.~Sanjanasri, and P.~Nedungadi, ``University students as early adopters of chatgpt: Innovation diffusion study,'' 2023.

\bibitem[Wang et~al.(2023)Wang, Vargas-Diaz, Brown, and Chen]{wang2023towards}
T.~Wang, D.~Vargas-Diaz, C.~Brown, and Y.~Chen, ``Towards adapting computer science courses to {AI} assistants' capabilities,'' \emph{arXiv preprint arXiv:2306.03289}, 2023.

\bibitem[Zastudil et~al.(2023)Zastudil, Rogalska, Kapp, Vaughn, and MacNeil]{zastudil2023generative}
C.~Zastudil, M.~Rogalska, C.~Kapp, J.~Vaughn, and S.~MacNeil, ``Generative ai in computing education: Perspectives of students and instructors,'' in \emph{2023 IEEE Frontiers in Education Conference (FIE)}.\hskip 1em plus 0.5em minus 0.4em\relax IEEE, 2023, pp. 1--9.

\bibitem[Barke et~al.(2023)Barke, James, and Polikarpova]{barke2023grounded}
S.~Barke, M.~B. James, and N.~Polikarpova, ``Grounded copilot: How programmers interact with code-generating models,'' \emph{Proceedings of the ACM on Programming Languages}, vol.~7, no. OOPSLA1, pp. 85--111, 2023.

\bibitem[Kazemitabaar et~al.(2023)Kazemitabaar, Chow, Ma, Ericson, Weintrop, and Grossman]{kazemitabaar2023studying}
M.~Kazemitabaar, J.~Chow, C.~K.~T. Ma, B.~J. Ericson, D.~Weintrop, and T.~Grossman, ``Studying the effect of {AI} code generators on supporting novice learners in introductory programming,'' in \emph{Proceedings of the 2023 CHI Conference on Human Factors in Computing Systems}, 2023, pp. 1--23.

\bibitem[Mozannar et~al.(2022)Mozannar, Bansal, Fourney, and Horvitz]{mozannar2022reading}
H.~Mozannar, G.~Bansal, A.~Fourney, and E.~Horvitz, ``Reading between the lines: Modeling user behavior and costs in ai-assisted programming,'' \emph{arXiv preprint arXiv:2210.14306}, 2022.

\bibitem[Ross et~al.(2023)Ross, Martinez, Houde, Muller, and Weisz]{ross2023programmer}
S.~I. Ross, F.~Martinez, S.~Houde, M.~Muller, and J.~D. Weisz, ``The programmer’s assistant: Conversational interaction with a large language model for software development,'' in \emph{Proceedings of the 28th International Conference on Intelligent User Interfaces}, 2023, pp. 491--514.

\bibitem[Vaithilingam et~al.(2022)Vaithilingam, Zhang, and Glassman]{vaithilingam2022expectation}
P.~Vaithilingam, T.~Zhang, and E.~L. Glassman, ``Expectation vs. experience: Evaluating the usability of code generation tools powered by large language models,'' in \emph{Chi conference on human factors in computing systems extended abstracts}, 2022, pp. 1--7.

\bibitem[Xu et~al.(2022)Xu, Vasilescu, and Neubig]{xu2022ide}
F.~F. Xu, B.~Vasilescu, and G.~Neubig, ``In-ide code generation from natural language: Promise and challenges,'' \emph{ACM Transactions on Software Engineering and Methodology (TOSEM)}, vol.~31, no.~2, pp. 1--47, 2022.

\bibitem[Ziegler et~al.(2022)Ziegler, Kalliamvakou, Li, Rice, Rifkin, Simister, Sittampalam, and Aftandilian]{ziegler2022productivity}
A.~Ziegler, E.~Kalliamvakou, X.~A. Li, A.~Rice, D.~Rifkin, S.~Simister, G.~Sittampalam, and E.~Aftandilian, ``Productivity assessment of neural code completion,'' in \emph{Proceedings of the 6th ACM SIGPLAN International Symposium on Machine Programming}, 2022, pp. 21--29.

\bibitem[Prather et~al.(2023{\natexlab{b}})Prather, Reeves, Denny, Becker, Leinonen, Luxton-Reilly, Powell, Finnie-Ansley, and Santos]{prather2023s}
J.~Prather, B.~N. Reeves, P.~Denny, B.~A. Becker, J.~Leinonen, A.~Luxton-Reilly, G.~Powell, J.~Finnie-Ansley, and E.~A. Santos, ``“it’s weird that it knows what i want”: Usability and interactions with copilot for novice programmers,'' \emph{ACM Transactions on Computer-Human Interaction}, vol.~31, no.~1, pp. 1--31, 2023.

\bibitem[Bird et~al.(2023)Bird, Ford, Zimmermann, Forsgren, Kalliamvakou, Lowdermilk, and Gazit]{bird2023taking}
C.~Bird, D.~Ford, T.~Zimmermann, N.~Forsgren, E.~Kalliamvakou, T.~Lowdermilk, and I.~Gazit, ``Taking flight with copilot,'' \emph{Communications of the ACM}, vol.~66, no.~6, pp. 56--62, 2023.

\bibitem[Amershi et~al.(2019)Amershi, Weld, Vorvoreanu, Fourney, Nushi, Collisson, Suh, Iqbal, Bennett, Inkpen, et~al.]{amershi2019guidelines}
S.~Amershi, D.~Weld, M.~Vorvoreanu, A.~Fourney, B.~Nushi, P.~Collisson, J.~Suh, S.~Iqbal, P.~N. Bennett, K.~Inkpen \emph{et~al.}, ``Guidelines for human-{AI} interaction,'' in \emph{Proceedings of the 2019 chi conference on human factors in computing systems}, 2019, pp. 1--13.

\bibitem[Prather et~al.(2024)Prather, Reeves, Leinonen, MacNeil, Randrianasolo, Becker, Kimmel, Wright, and Briggs]{prather2024widening}
J.~Prather, B.~N. Reeves, J.~Leinonen, S.~MacNeil, A.~S. Randrianasolo, B.~A. Becker, B.~Kimmel, J.~Wright, and B.~Briggs, ``The widening gap: The benefits and harms of generative {AI} for novice programmers,'' in \emph{Proceedings of the 2024 ACM Conference on International Computing Education Research-Volume 1}, 2024, pp. 469--486.

\bibitem[Skripchuk et~al.(2024)Skripchuk, Bacher, and Price]{skripchuk2024investigation}
J.~Skripchuk, J.~Bacher, and T.~Price, ``An investigation of the drivers of novice programmers' intentions to use web search and {GenAI},'' in \emph{Proceedings of the 2024 ACM Conference on International Computing Education Research-Volume 1}, 2024, pp. 487--501.

\bibitem[Ernst and Bavota(2022)]{ernst2022ai}
N.~A. Ernst and G.~Bavota, ``{AI}-driven development is here: Should you worry?'' \emph{IEEE Software}, vol.~39, no.~2, pp. 106--110, 2022.

\bibitem[Becker et~al.(2019)Becker, Denny, Pettit, Bouchard, Bouvier, Harrington, Kamil, Karkare, McDonald, Osera, et~al.]{becker2019compiler}
B.~A. Becker, P.~Denny, R.~Pettit, D.~Bouchard, D.~J. Bouvier, B.~Harrington, A.~Kamil, A.~Karkare, C.~McDonald, P.-M. Osera \emph{et~al.}, ``Compiler error messages considered unhelpful: The landscape of text-based programming error message research,'' \emph{Proceedings of the working group reports on innovation and technology in computer science education}, pp. 177--210, 2019.

\bibitem[Finnie-Ansley et~al.(2022)Finnie-Ansley, Denny, Becker, Luxton-Reilly, and Prather]{finnie2022robots}
J.~Finnie-Ansley, P.~Denny, B.~A. Becker, A.~Luxton-Reilly, and J.~Prather, ``The robots are coming: Exploring the implications of openai codex on introductory programming,'' in \emph{Proceedings of the 24th Australasian Computing Education Conference}, 2022, pp. 10--19.

\bibitem[Sarsa et~al.(2022)Sarsa, Denny, Hellas, and Leinonen]{sarsa2022automatic}
S.~Sarsa, P.~Denny, A.~Hellas, and J.~Leinonen, ``Automatic generation of programming exercises and code explanations using large language models,'' in \emph{Proceedings of the 2022 ACM Conference on International Computing Education Research-Volume 1}, 2022, pp. 27--43.

\bibitem[Savelka et~al.(2023)Savelka, Agarwal, Bogart, Song, and Sakr]{savelka2023can}
J.~Savelka, A.~Agarwal, C.~Bogart, Y.~Song, and M.~Sakr, ``Can generative pre-trained transformers (gpt) pass assessments in higher education programming courses?'' in \emph{Proceedings of the 2023 Conference on Innovation and Technology in Computer Science Education V. 1}, 2023, pp. 117--123.

\bibitem[Leinonen et~al.(2023{\natexlab{a}})Leinonen, Denny, MacNeil, Sarsa, Bernstein, Kim, Tran, and Hellas]{leinonen2023comparing}
J.~Leinonen, P.~Denny, S.~MacNeil, S.~Sarsa, S.~Bernstein, J.~Kim, A.~Tran, and A.~Hellas, ``Comparing code explanations created by students and large language models,'' in \emph{Proceedings of the 2023 Conference on Innovation and Technology in Computer Science Education V. 1}, 2023, pp. 124--130.

\bibitem[MacNeil et~al.(2023)MacNeil, Tran, Hellas, Kim, Sarsa, Denny, Bernstein, and Leinonen]{macneil2023experiences}
S.~MacNeil, A.~Tran, A.~Hellas, J.~Kim, S.~Sarsa, P.~Denny, S.~Bernstein, and J.~Leinonen, ``Experiences from using code explanations generated by large language models in a web software development e-book,'' in \emph{Proceedings of the 54th ACM Technical Symposium on Computer Science Education V. 1}, 2023, pp. 931--937.

\bibitem[Wermelinger(2023)]{wermelinger2023using}
M.~Wermelinger, ``Using {GitHub} copilot to solve simple programming problems,'' in \emph{Proceedings of the 54th ACM Technical Symposium on Computer Science Education V. 1}, 2023, pp. 172--178.

\bibitem[Leinonen et~al.(2023{\natexlab{b}})Leinonen, Hellas, Sarsa, Reeves, Denny, Prather, and Becker]{leinonen2023using}
J.~Leinonen, A.~Hellas, S.~Sarsa, B.~Reeves, P.~Denny, J.~Prather, and B.~A. Becker, ``Using large language models to enhance programming error messages,'' in \emph{Proceedings of the 54th ACM Technical Symposium on Computer Science Education V. 1}, 2023, pp. 563--569.

\bibitem[Chan and Lee(2023)]{chan2023ai}
C.~K.~Y. Chan and K.~K. Lee, ``The {AI} generation gap: Are {Gen Z} students more interested in adopting generative {AI} such as chatgpt in teaching and learning than their {Gen X} and millennial generation teachers?'' \emph{Smart Learning Environments}, vol.~10, no.~1, p.~60, 2023.

\bibitem[Padiyath et~al.(2024)Padiyath, Hou, Pang, Viramontes~Vargas, Gu, Nelson-Fromm, Wu, Guzdial, and Ericson]{padiyath2024insights}
A.~Padiyath, X.~Hou, A.~Pang, D.~Viramontes~Vargas, X.~Gu, T.~Nelson-Fromm, Z.~Wu, M.~Guzdial, and B.~Ericson, ``Insights from social shaping theory: The appropriation of large language models in an undergraduate programming course,'' in \emph{Proceedings of the 2024 ACM Conference on International Computing Education Research-Volume 1}, 2024, pp. 114--130.

\bibitem[Kim et~al.(2024)Kim, Ko, Myers, and Bach]{kim2024chatgpt}
N.~W. Kim, H.-K. Ko, G.~Myers, and B.~Bach, ``Chatgpt in data visualization education: A student perspective,'' \emph{arXiv preprint arXiv:2405.00748}, 2024.

\bibitem[Anonymous(2024)]{supplemental2023}
\BIBentryALTinterwordspacing
Anonymous, ``{Supplemental Material for GenAI Utilization Among SE Students},'' Anonymous, October 2024. [Online]. Available: \url{https://doi.org/10.5281/zenodo.13882829}
\BIBentrySTDinterwordspacing

\bibitem[Krathwohl(2002)]{krathwohl2002revision}
D.~R. Krathwohl, ``A revision of bloom's taxonomy: An overview,'' \emph{Theory into practice}, vol.~41, no.~4, pp. 212--218, 2002.

\bibitem[Forehand et~al.(2005)]{forehand2005bloom}
M.~Forehand \emph{et~al.}, ``Bloom's taxonomy: Original and revised,'' \emph{Emerging perspectives on learning, teaching, and technology}, vol.~8, pp. 41--44, 2005.

\bibitem[Cabezas et~al.(2020)Cabezas, Segovia, Caratozzolo, and Webb]{cabezas2020using}
I.~Cabezas, R.~Segovia, P.~Caratozzolo, and E.~Webb, ``Using software engineering design principles as tools for freshman students learning,'' in \emph{2020 IEEE Frontiers in Education Conference (FIE)}.\hskip 1em plus 0.5em minus 0.4em\relax IEEE, 2020, pp. 1--5.

\bibitem[Bargh and Chartrand(2000)]{bargh2000mind}
J.~A. Bargh and T.~L. Chartrand, ``The mind in the middle,'' \emph{Handbook of research methods in social and personality psychology}, vol.~2, pp. 253--285, 2000.

\bibitem[Furnham and Boo(2011)]{furnham2011literature}
A.~Furnham and H.~C. Boo, ``A literature review of the anchoring effect,'' \emph{The journal of socio-economics}, vol.~40, no.~1, pp. 35--42, 2011.

\bibitem[Creswell and Poth(2016)]{creswell2016qualitative}
J.~W. Creswell and C.~N. Poth, \emph{Qualitative inquiry and research design: Choosing among five approaches}.\hskip 1em plus 0.5em minus 0.4em\relax Sage publications, 2016.

\bibitem[Braun and Clark(2006)]{braun2006using}
V.~Braun and V.~Clark, ``Using thematic analysis in psychology,'' \emph{Qualitative research in psychology}, vol.~3, no.~2, pp. 77--101, 2006.

\bibitem[Braun and Clarke(2022)]{braun2022conceptual}
V.~Braun and V.~Clarke, ``Conceptual and design thinking for thematic analysis.'' \emph{Qualitative Psychology}, vol.~9, no.~1, p.~3, 2022.

\bibitem[atl(2024)]{atlas_website}
\BIBentryALTinterwordspacing
``Atlas.ti (version 3) [computer software],'' 2024, accessed: 2024-09-07. [Online]. Available: \url{https://atlasti.com/}
\BIBentrySTDinterwordspacing

\bibitem[McDonald et~al.(2019)McDonald, Schoenebeck, and Forte]{mcdonald2019reliability}
N.~McDonald, S.~Schoenebeck, and A.~Forte, ``Reliability and inter-rater reliability in qualitative research: Norms and guidelines for {CSCW} and {HCI} practice,'' \emph{Proceedings of the ACM on human-computer interaction}, vol.~3, no. CSCW, pp. 1--23, 2019.

\bibitem[Francis et~al.(2010)Francis, Johnston, Robertson, Glidewell, Entwistle, Eccles, and Grimshaw]{francis2010adequate}
J.~J. Francis, M.~Johnston, C.~Robertson, L.~Glidewell, V.~Entwistle, M.~P. Eccles, and J.~M. Grimshaw, ``What is an adequate sample size? operationalising data saturation for theory-based interview studies,'' \emph{Psychology and health}, vol.~25, no.~10, pp. 1229--1245, 2010.

\bibitem[Biswas(2023)]{biswas2023evaluating}
S.~Biswas, ``Evaluating errors and improving performance of chatgpt,'' \emph{International Journal of Clinical and Medical Education Research}, vol.~2, no.~6, pp. 182--188, 2023.

\bibitem[Plass et~al.(2010)Plass, Moreno, and Br{\"u}nken]{plass2010cognitive}
J.~L. Plass, R.~Moreno, and R.~Br{\"u}nken, ``Cognitive load theory,'' 2010.

\bibitem[Bandura et~al.(1986)]{bandura1986social}
A.~Bandura \emph{et~al.}, ``Social foundations of thought and action,'' \emph{Englewood Cliffs, NJ}, vol. 1986, no. 23-28, p.~2, 1986.

\bibitem[Deci and Ryan(2012)]{deci2012self}
E.~L. Deci and R.~M. Ryan, ``Self-determination theory,'' \emph{Handbook of theories of social psychology}, vol.~1, no.~20, pp. 416--436, 2012.

\bibitem[Davis et~al.(1989)Davis, Bagozzi, and Warshaw]{davis1989technology}
F.~D. Davis, R.~Bagozzi, and P.~Warshaw, ``Technology acceptance model,'' \emph{J Manag Sci}, vol.~35, no.~8, pp. 982--1003, 1989.

\bibitem[{Harvard Office of Undergraduate Education}(2023)]{Harvard_OUE_2023}
{Harvard Office of Undergraduate Education}, ``{AI Guidance \& FAQs},'' \url{https://oue.fas.harvard.edu/ai-guidance-faqs}, 2023.

\bibitem[{UCLA Center for the Advancement of Teaching}(2023)]{UCLA_CAT_2023}
{UCLA Center for the Advancement of Teaching}, ``Guidance for the use of {Generative AI},'' \url{https://teaching.ucla.edu/resources/guidance-for-the-use-of-generative-ai/}, 2023.

\bibitem[Amoozadeh et~al.(2024)Amoozadeh, Daniels, Nam, Kumar, Chen, Hilton, Srinivasa~Ragavan, and Alipour]{amoozadeh2024trust}
M.~Amoozadeh, D.~Daniels, D.~Nam, A.~Kumar, S.~Chen, M.~Hilton, S.~Srinivasa~Ragavan, and M.~A. Alipour, ``Trust in {Generative AI} among students: An exploratory study,'' in \emph{55th ACM Technical Symposium on Computer Science Education V. 1}, 2024, pp. 67--73.

\bibitem[Choudhuri et~al.(2024{\natexlab{b}})Choudhuri, Trinkenreich, Pandita, Kalliamvakou, Steinmacher, Gerosa, Sanchez, and Sarma]{choudhuri2024guides}
R.~Choudhuri, B.~Trinkenreich, R.~Pandita, E.~Kalliamvakou, I.~Steinmacher, M.~Gerosa, C.~Sanchez, and A.~Sarma, ``{What Guides Our Choices? Modeling Developers' Trust and Behavioral Intentions Towards GenAI},'' \emph{arXiv preprint arXiv:2409.04099}, 2024.

\bibitem[Bird et~al.(2022)Bird, Ford, Zimmermann, Forsgren, Kalliamvakou, Lowdermilk, and Gazit]{bird2022taking}
C.~Bird, D.~Ford, T.~Zimmermann, N.~Forsgren, E.~Kalliamvakou, T.~Lowdermilk, and I.~Gazit, ``Taking flight with copilot: Early insights and opportunities of ai-powered pair-programming tools,'' \emph{Queue}, vol.~20, no.~6, pp. 35--57, 2022.

\bibitem[White et~al.(2023)White, Fu, Hays, Sandborn, Olea, Gilbert, Elnashar, Spencer-Smith, and Schmidt]{white2023prompt}
J.~White, Q.~Fu, S.~Hays, M.~Sandborn, C.~Olea, H.~Gilbert, A.~Elnashar, J.~Spencer-Smith, and D.~C. Schmidt, ``A prompt pattern catalog to enhance prompt engineering with chatgpt,'' \emph{arXiv preprint arXiv:2302.11382}, 2023.

\bibitem[Merriam and Tisdell(2015)]{merriam2015qualitative}
S.~B. Merriam and E.~J. Tisdell, \emph{Qualitative research: A guide to design and implementation}.\hskip 1em plus 0.5em minus 0.4em\relax John Wiley \& Sons, 2015.

\bibitem[Denzin and Lincoln(2011)]{denzin2011sage}
N.~K. Denzin and Y.~S. Lincoln, \emph{The Sage handbook of qualitative research}.\hskip 1em plus 0.5em minus 0.4em\relax sage, 2011.

\end{thebibliography}

\end{document}